\documentclass[10pt]{article}
\newtheorem{theorem}{Theorem}[section]
\newtheorem{lemma}[theorem]{Lemma}
\newtheorem{remark}[theorem]{Remark}
\newtheorem{definition}[theorem]{Definition}
\newtheorem{corollary}[theorem]{Corollary}

\usepackage{amsmath}
\usepackage{amsfonts}
\usepackage{amstext}
\usepackage{amssymb}
\usepackage{setspace}

\begin{document}
\title{Asymptotic Behavior of the Stock Price Distribution Density and Implied Volatility in Stochastic Volatility Models}
\date{}
\author{Archil Gulisashvili $\cdot$ \\
Elias M. Stein}
\maketitle
\vspace{0.2in}
\bf Abstract \rm We study the asymptotic behavior of distribution densities arising in stock price models 
with stochastic volatility.
The main objects of our interest in the present paper are the density of time averages of the squared 
volatility process and the density of the stock price process in the Stein-Stein and the Heston model. We find explicit formulas for 
leading terms in asymptotic expansions of these densities and give error estimates. As an application of our results, sharp asymptotic formulas for the implied volatility in the Stein-Stein and the Heston model are obtained. 
\\
\\
\bf Keywords\,\, \rm Stein-Stein model $\cdot$ Heston model $\cdot$ Mixing distribution density $\cdot$ Stock price $\cdot$
Bessel processes $\cdot$ Ornstein-Uhlenbeck processes $\cdot$ CIR processes $\cdot$ Asymptotic formulas $\cdot$ Implied volatility
\hspace{1in} 
\section{Introduction}\label{S:ashes}
\normalsize
In \cite{keyGS1} and \cite{keyGS2}, we found sharp asymptotic formulas 
for the distribution density of the stock price process in the Hull-White model. These formulas were used in \cite{keyGS3} to characterize
the asymptotic behavior of the implied volatility in the Hull-White model. The present paper is a continuation of \cite{keyGS1}, \cite{keyGS2}, 
and \cite{keyGS3}. It concerns the asymptotic behavior of various distribution densities arising in the 
Stein-Stein and the Heston model.
\footnotetext{A. Gulisashvili \\
Department of Mathematics, Ohio University, Athens, OH 45701, USA \\
e-mail: guli@math.ohiou.edu \\
\\
E. M. Stein \\
Department of Mathematics, Princeton University, Princeton, NJ 08540, USA \\
e-mail: stein@math.princeton.edu}

The stochastic differential equations characterizing the stock price process $X_t$ and the volatility process $Y_t$ in the Stein-Stein model have the following form:
\begin{equation}
\left\{\begin{array}{ll}
dX_t=\mu X_tdt+\left|Y_t\right|X_tdW_t \\
dY_t=q\left(m-Y_t\right)dt+\sigma dZ_t,
\end{array}
\right.
\label{E:SS}
\end{equation}
where $W_t$ and $Z_t$ are standard Brownian motions. The initial condition for $X_t$ is denoted by $x_0$ and for $Y_t$ by $y_0$. The stochastic volatility model in (\ref{E:SS}) was introduced and studied in \cite{keySS1}. In this model, the absolute value of an Ornstein-Uhlenbeck 
process plays the role of the volatility of the stock. We will assume that
that $\mu\in\mathbb{R}$, $q> 0$, $m\ge 0$, and $\sigma> 0$.

The Heston model was introduced in \cite{keyH}. In this model, the stock price process $X_t$ and the volatility process $Y_t$ satisfy the
following system of stochastic differential equations:
\begin{equation}
\left\{\begin{array}{ll}
dX_t=\mu X_tdt+\sqrt{Y_t}X_tdW_t \\
dY_t=\left(a+bY_t\right)dt+c\sqrt{Y_t}dZ_t,
\end{array}
\right.
\label{E:H}
\end{equation}
where $W_t$ and $Z_t$ are standard Brownian motions. We will assume that $\mu\in\mathbb{R}$, $a\ge 0$, $b< 0$, and $c> 0$. 
The initial conditions for $X_t$ and $Y_t$ are denoted by $x_0$ and $y_0$, respectively. The volatility equation in (\ref{E:H}) can be rewritten in the mean-reverting form. This gives
$$
\left\{\begin{array}{ll}
dX_t=\mu X_tdt+\sqrt{Y_t}X_tdW_t \\
dY_t=r\left(m-Y_t\right)dt+c\sqrt{Y_t}dZ_t,
\end{array}
\right.
$$
where $r=-b$ and $m=-\frac{a}{b}$. The volatility equation in (\ref{E:H}) is uniquely solvable in the strong sense, and the solution $Y_t$ is a positive stochastic process. This process is called a Cox-Ingersoll-Ross process (a CIR-process). This process was studied in \cite{keyCIR}. Interesting results concerning the Heston model were obtained in \cite{keyDY}.

It will be assumed throughout the present paper that the models described by (\ref{E:SS}) and (\ref{E:H}) are uncorrelated. This means that the 
Brownian motions $W_t$ and $Z_t$ driving the stock price equation and the volatility equation in (\ref{E:SS}) and (\ref{E:H}) are independent. 
In the analysis of the probability distribution of the stock price $X_t$, the mean-square averages of the volatility process over finite time intervals play an important role. For the Stein-Stein model, we set 
\begin{equation}
\alpha_t=\left\{\frac{1}{t}\int_0^tY_s^2ds\right\}^{\frac{1}{2}}.
\label{E:mix1}
\end{equation}
Here $Y_t$ satisfies the second equation in (\ref{E:SS}). For the Heston model, we have
\begin{equation}
\alpha_t=\left\{\frac{1}{t}\int_0^tY_sds\right\}^{\frac{1}{2}},
\label{E:mix2}
\end{equation}
where $Y_t$ satisfies the second equation in (\ref{E:H}). It will be shown below that for every 
$t> 0$, the probability distribution of the random variable $\alpha_t$ defined by (\ref{E:mix1}) for the model in (\ref{E:SS}) and by (\ref{E:mix2})
for the model in (\ref{E:H}) 
admits a distribution density $m_t$ (see Lemma \ref{L:disden} and Remark \ref{R:mixis} below). The function $m_t$ is called the mixing distribution density. 

The distribution density of the stock price $X_t$ will be denoted by $D_t$. The existence of the density $D_t$ follows from formula (\ref{E:stvoll1}). 
In this paper, we obtain sharp asymptotic formulas for the distribution density $D_t$ in the case of the uncorrelated Stein-Stein and Heston models. Note that in \cite{keyDY} and \cite{keySS1}, the behavior of $D_t$ was studied for the Heston model and the Stein-Stein model, respectively, using a rough logarithmic scale in the asymptotic formulas. Moreover, no error estimates were given in these papers. The results established in the present paper are considerably sharper. We find explicit formulas for leading terms in asymptotic expansions of $D_t$ in the Heston and the Stein-Stein model with error estimates. It would be interesting to obtain similar results for correlated models, since such models have more applications in finance. We hope that the methods employed in the present paper may be useful in the study of the correlated case.

We will next quickly overview the structure of the present paper. In Section \ref{S:afsp}, the main results of the paper (theorems \ref{T:maino}, 
\ref{T:main}, \ref{T:ohh}, and \ref{T:oh}) are formulated. They concern the asymptotic behavior 
of the mixing distribution density $m_t$ and the stock price distribution density $D_t$ in the 
Heston model and the Stein-Stein model. Section \ref{S:imvol} is devoted to applications of our main results. In this section, we obtain sharp asymptotic formulas for the implied volatility in the Heston and the Stein-Stein model. In section \ref{S:CIRB}, we gather several known facts about CIR-processes and Bessel processes. We also formulate a theorem of Pitman and Yor concerning exponential functionals of Bessel processes. This theorem plays an important role in the present paper. In Section \ref{S:ilt}, we prove an asymptotic inversion theorem for the Laplace transform in a certain class of functions (see Theorem \ref{T:lti}). This theorem is useful in the study of the asymptotic behavior of the mixing distribution density, since it is often easier to find explicit formulas for the Laplace transform of this density than to characterize the density itself. 
In sections \ref{S:firh2} - \ref{S:qv}, we prove theorems \ref{T:maino} - \ref{T:oh} and describe 
the constants appearing in these theorems. 
\section{Asymptotic formulas for distribution densities}\label{S:afsp}
The next four theorems are the main results of the present paper. The first two of them provide explicit formulas for the leading term in the asymptotic expansion of the distribution density $D_t$ of the stock price $X_t$ in the Stein-Stein model and the Heston model, while the other two concern the asymptotic behavior of the mixing distribution density $m_t$ in these models.
\begin{theorem}\label{T:maino}
Let $D_t$ be the stock price distribution density in model (\ref{E:SS}) with $q\ge 0$, $m\ge 0$, and $\sigma> 0$.
Then there exist positive constants $B_1$,
$B_2$, and $B_3$ such that 
\begin{equation}
D_t\left(x_0e^{\mu t}x\right)=B_1x^{-B_3}e^{B_2\sqrt{\log x}}(\log x)^{-\frac{1}{2}}
\left(1+O\left((\log x)^{-\frac{1}{4}}\right)\right),\quad x\rightarrow\infty.
\label{E:dop1}
\end{equation}
\end{theorem} 
\begin{theorem}\label{T:main}
Let $D_t$ be the stock price distribution density in model (\ref{E:H}) with $a\ge 0$, $b\le 0$, and $c> 0$. Then there exist positive constants $A_1$,
$A_2$, and $A_3$ such that 
\begin{equation}
D_t\left(x_0e^{\mu t}x\right)=A_1x^{-A_3}e^{A_2\sqrt{\log x}}(\log x)^{-\frac{3}{4}+\frac{a}{c^2}}
\left(1+O\left((\log x)^{-\frac{1}{4}}\right)\right),\quad x\rightarrow\infty.
\label{E:dopo}
\end{equation}
\end{theorem}
\begin{theorem}\label{T:ohh}
Suppose that $q\ge 0$, $m\ge 0$, and $\sigma> 0$ in model (\ref{E:SS}). Then there exist positive constants $E$, $F$, and $G$ such that
\begin{equation}
m_t(y)=Ee^{-Gy^2}e^{Fy}
\left(1+O\left(y^{-\frac{1}{2}}\right)\right),\quad y\rightarrow\infty.
\label{E:ohh1}
\end{equation}
\end{theorem}
\begin{theorem}\label{T:oh}
Suppose that $a\ge 0$, $b\le 0$, and $c> 0$ in model (\ref{E:H}). Then there exist positive constants $A$, $B$, and $C$ such that
\begin{equation}
m_t(y)=Ae^{-Cy^2}e^{By}y^{-\frac{1}{2}+\frac{2a}{c^2}}
\left(1+O\left(y^{-\frac{1}{2}}\right)\right),\quad y\rightarrow\infty.
\label{E:oh1}
\end{equation}
\end{theorem}

It is known that for uncorrelated stochastic volatility models, the asymptotic behavior of the stock price distribution density $D_t$ near zero is determined by the behavior of $D_t$ near infinity. Indeed, we have
$\displaystyle{D_t\left(x_0e^{\mu t}x^{-1}\right)=x^3D_t\left(x_0e^{\mu t}x\right)}$, $x> 0$. This equality can be derived from the formula
\begin{equation}
 D_t\left(x_0e^{\mu t}x\right)=\frac{1}{x_0e^{\mu t}\sqrt{2\pi t}}x^{-\frac{3}{2}}\int_0^{\infty}y^{-1}m_t(y)
\exp\left\{-\left[\frac{1}{2ty^2}\log^2x+\frac{ty^2}{8}\right]\right\}dy
\label{E:stvoll1}
\end{equation}
(see, e.g., Section 4 in \cite{keyGS2}). It follows from Theorem \ref{T:main} that the stock price distribution density $D_t(x)$ in the Heston model behaves at infinity roughly as the function
$x^{-A_3}$ and at zero as the function $x^{A_3-3}$. For the Stein-Stein model, Theorem \ref{T:maino} implies that $D_t(x)$ behaves at 
infinity roughly as the function $x^{-B_3}$ and at zero as the function $x^{B_3-3}$. In \cite{keySS1}, the latter fact was established (in part, only heuristically) in the logarithmic scale 
(the same scale was used in \cite{keyDY}). More precisely, the following definition of the asymptotic 
equivalence was used in \cite{keyDY} and \cite{keySS1}: Two functions $F(x)$ and $G(x)$ are called 
asymptotically equivalent as $x\rightarrow\infty$ (or as $x\rightarrow 0$) if 
$\frac{\log F(x)}{\log G(x)}\rightarrow 1\quad\mbox{as}\quad x\rightarrow\infty\quad\mbox{(or $x\rightarrow 0$)}$.
It is not hard to see that the constant $\gamma$ in \cite{keySS1}, wich characterizes the decay of the stock price density 
$D_t$ near infinity (see formula (20) in \cite{keySS1}), coincides with the constant 
$B_3$ in (\ref{E:dop1}) (see lemmas \ref{L:con1} and \ref{L:con2} for the description of the constant $B_3$). However, there is an error in formula (21) in \cite{keySS1}. It is stated that $D_t(x)$ behaves near zero as the function $x^{-1+\gamma}$. By Theorem \ref{T:maino}, the correct power function that characterizes 
the behavior of the density $D_t$ near zero is the function $x^{-3+\gamma}$. 

The values of the constants $A_3$ and $B_3$ appearing in theorems \ref{T:maino} and \ref{T:main} 
are given by the following formulas: 
$$
A_3=A_3(t,b,c)=\frac{3}{2}+\frac{\sqrt{8C+t}}{2\sqrt{t}},\quad C=C(t,b,c)
=\frac{t}{2c^2}\left(b^2+\frac{4}{t^2}r^2_{\frac{t|b|}{2}}\right),
$$
and
$$
B_3=B_3(t,b,c)=\frac{3}{2}+\frac{\sqrt{8G+t}}{2\sqrt{t}},\quad G=G(t,q,\sigma)
=\frac{t}{2\sigma^2}\left(q^2+\frac{1}{t^2}r^2_{qt}\right)
$$
(see lemmas \ref{L:con1}, \ref{L:con2}, \ref{L:con3}, and \ref{L:con4} below). Here 
$r_s$ denotes the smallest positive root of the entire function $z\mapsto z\,cos z+s\,sin z$. 
The zeroes of this function are studied in Section \ref{S:firh2}. Note that $A_3$ does not depend on $a$, while $B_3$ does 
not depend on $m$. It is clear that $A_3> 2$ and $B_3> 2$. In \cite{keyGS1} and \cite{keyGS2}, we studied the tail behavior of the stock price distribution density 
$D_t$ in the Hull-White model. It was established that the function $D_t(x)$ behaves at infinitiy like the function $x^{-2}$ on the power function scale. This is an extremely slow behavior. No uncorrelated stochastic volatility model has the function $D_t(x)$ decaying like 
$x^{-2+\epsilon}$, $\epsilon> 0$ (see \cite{keyGS2}). The tail of the stock price distribution in the Hull-White model is ``fatter"
than the corresponding tail in the Heston and the Stein-Stein model.
\section{Applications. Asymptotic behavior of the implied volatility}\label{S:imvol}
The implied volatility in an option pricing model is the volatility in the Black-Scholes model such that 
the corresponding Black-Scholes price of the option is equal to its price in the model under consideration. 
In the present paper, we will only consider European call options, and the implied volatility will be studied 
as a function of the strike price $K$. Let us denote by $V_0$ the pricing function for the European call option in 
the Heston (or the Stein-Stein) model. Then 
the impled volatility $I(K)$ satisfies $C_{BS}(K,I(K))=V_0(K)$, where $C_{BS}$ stands for the Black-Scholes pricing function. 

In \cite{keyGS3}, we studied the implied volatility in the Hull-White model. In the present paper, we 
characterize the asymptotic behavior of the implied volatility in the Stein-Stein and the Heston model. Our asymptotic formulas are sharper 
than the formulas which can be obtained form more general results due to Lee, Benaim, and Friz (see \cite{keyL}, \cite{keyBF}, \cite{keyBFL}).
We will first consider the model in (\ref{E:H}). It will be assumed below that the market price of volatility risk $\gamma$ is equal 
to zero (see, e.g., \cite{keyFPS} for the definition of $\gamma$). Then, 
under the corresponding martingale measure 
$\mathbb{P}^{*}$, the system of stochastic differential equations in (\ref{E:H}) can be rewritten in the following form:
\begin{equation}
\left\{\begin{array}{l}
dX_t=rX_tdt+Y_tX_tdW_t^{*} \\
dY_t=\left(a+bY_t\right)dt+c\sqrt{Y_t}dZ_t^{*},
\end{array}
\right.
\label{E:sde2}
\end{equation}
where $W_t^{*}$ and $Z_t^{*}$ are independent standard one-dimensional Brownian motions, 
and $r> 0$ is a constant interest rate. This means that under the measure $\mathbb{P}^{*}$ the new system also
describes a Heston model. This is the reason why we assumed that $\gamma=0$.

Let us consider a European call option associated with the stock price model in (\ref{E:sde2}). The price of such an option 
at $t=0$ is given by the following formula:
\begin{equation}
V_0(K)=\mathbb{E}^{*}\left[e^{-rT}\left(X_T-K\right)_{+}\right],
\label{E:opt}
\end{equation}
where $T$ is the expiration date and $K$ is the strike price. Put 
$D_T(x)=D_T\left(x;r,\nu,\xi,x_0,y_0\right)$. Then we have
\begin{equation}
V_0(K)=e^{-rT}\int_K^{\infty}xD_T(x)dx-e^{-rT}K\int_K^{\infty}
D_T(x)dx.
\label{E:option} 
\end{equation}

The implied volatility is often considered as a function of the log-strike $k$, which in our case is related to the strike price $K$ by the formula
$k=\log\frac{K}{x_0e^{rT}}$. In terms of $k$, the implied volatility is defined as follows:
$\hat{I}(k)=I(K)$, $-\infty< k<\infty$, $0< K<\infty$. For uncorrelated stochastic volatlity models, the behavior of the implied volatility near zero is completely determined by how it behaves near infinity. 
More precisely, it is known that the implied volatility is symmetric in the following sense:
$I\left(\left(x_0e^{rT}\right)^2K^{-1}\right)=I\left(K\right)$
for all $K> 0$ (see, e.g., \cite{keyGS2}). It is clear that the previous equality can be formulated in terms of the log-strike $k$ as follows:
$\hat{I}(k)=\hat{I}(-k)$, $-\infty< k<\infty$.
 
Important results concerning the behavior of the implied volatility $\hat{I}(k)$ in a general case were obtained by Lee (see \cite{keyL}). He characterized the asymptotic behavior of the implied volatlity for large strikes in terms of the moments of the
stock price process. In \cite{keyBF}, the asymptotic behavior of the implied volatility was linked to the tail behavior of the distribution of the stock price process (see also \cite{keyBFL}). The reader interested in more aspects of the asymptotic behavior 
of the implied volatility can consult Chapter 5 in \cite{keyFPS}.

The following theorem will be established below:
\begin{theorem}\label{T:impik}
For the Heston model,
\begin{equation}
\hat{I}(k)=\beta_1k^{\frac{1}{2}}+\beta_2+\beta_3\frac{\log k}{k^{\frac{1}{2}}}+O\left(\frac{\psi(k)}{k^{\frac{1}{2}}}\right)
\label{E:impik1}
\end{equation}
as $k\rightarrow\infty$, where
$$
\beta_1=\frac{\sqrt{2}}{\sqrt{T}}\left(\sqrt{A_3-1}-\sqrt{A_3-2}\right),
$$
$$
\beta_2=\frac{A_2}{\sqrt{2T}}\left(\frac{1}{\sqrt{A_3-2}}-\frac{1}{\sqrt{A_3-1}}\right),
$$
$$
\beta_3=\frac{1}{\sqrt{2T}}\left(\frac{1}{4}-\frac{a}{c^2}\right)\left(\frac{1}{\sqrt{A_3-1}}-\frac{1}{\sqrt{A_3-2}}\right),
$$
and $\psi$ can be any positive increasing function on $(0,\infty)$ such that $\displaystyle{\lim_{k\rightarrow\infty}\psi(k)=\infty}$.

For the Stein-Stein model,
\begin{equation}
\hat{I}(k)=\gamma_1k^{\frac{1}{2}}+\gamma_2+O\left(\frac{\psi(k)}{k^{\frac{1}{2}}}\right)
\label{E:impik2}
\end{equation}
as $k\rightarrow\infty$, where
$$
\gamma_1=\frac{\sqrt{2}}{\sqrt{T}}\left(\sqrt{B_3-1}-\sqrt{B_3-2}\right),
$$
$$
\gamma_2=\frac{B_2}{\sqrt{2T}}\left(\frac{1}{\sqrt{B_3-2}}-\frac{1}{\sqrt{B_3-1}}\right),
$$
and the function $\psi$ is as above.
\end{theorem}
\section{CIR processes and Bessel processes}\label{S:CIRB}
The volatility in model (\ref{E:H}) is described by the square root of an CIR-process. It is known that CIR-processes are related to squared 
Bessel processes. In the present section, we gather several results concerning Bessel processes and CIR-processes.
Let $\delta\ge 0$ and $x\ge 0$, and consider the following stochastic differential equation:
$dT_t=\delta dt+2\sqrt{T_t}dZ_t$, $T_0=x$ a.s.
This equation has a unique nonnegative strong solution 
$T_t$, which is called the squared $\delta$-dimensional Bessel process started at $x$. The following notation is often used for the squared Bessel process: $T_t=BESQ^{\delta}_x(t)$. We refer the reader to \cite{keyBS,keyCS,keyD,keyGY,keyPY,keyRY} for more information on Bessel processes.

The next lemma links Bessel processes and CIR-processes (see, e.g., \cite{keyGY}).
\begin{lemma}\label{L:sviaz1}
Let $Y_t$ be a CIR process satisfying the equation
$dY_t=\left(a+bY_t\right)dt+c\sqrt{Y_t}dZ_t$ with $Y_0=x$ $\mathbb{P}$-a.s.,
and put $T_t=BESQ^{\frac{4a}{c^2}}_x(t)$. Then
$Y_t=e^{bt}T\left(\frac{c^2}{4b}\left(1-e^{-bt}\right)\right)$.
\end{lemma}
\begin{remark}\label{R:vto} \rm
If $b=0$, then we have 
$Y_t=BESQ_x^{\frac{4a}{c^2}}\left(\frac{c^2}{4}t\right)$.
\end{remark}

Pitman and Yor proved the following assertion (see \cite{keyPY,keyRY}):
\begin{theorem}\label{T:kri}
Let $\lambda> 0$. Then 
$$
\mathbb{E}\left[\exp\left\{-\frac{\lambda^2}{2}\int_0^tBESQ_x^{\delta}(u)du\right\}\right]
=\left[\cosh(\lambda t)\right]^{-\frac{\delta}{2}}\exp\left\{-\frac{x\lambda}{2}\tanh(\lambda t)
\right\}.
$$
\end{theorem}
The next statement can be derived from Theorem \ref{T:kri}. This statement can be found, e.g., in \cite{keyBS}.
\begin{theorem}\label{T:liu}
Let $a\ge 0$, $b< 0$, $c> 0$, and let $Y_t$ be a CIR-process in Lemma \ref{L:sviaz1} such that $Y_0=y_0$ a.s. Then
for every $\eta>\frac{1}{2}$,
\begin{align*}
& \\
&=\exp\left\{-\frac{abt}{c^2}\right\}\left(\frac{2\eta}
{2\eta\cosh(bt\eta)-\sinh(bt\eta)}\right)^{\frac{2a}{c^2}}\exp\left\{-\frac{by_0\left(4\eta^2-1\right)\sinh(bt\eta)}
{2c^2\eta\cosh(bt\eta)-c^2\sinh(bt\eta)}\right\}.
\end{align*}
\end{theorem}
Theorem \ref{T:liu} is equivalent to the following assertion:
\begin{theorem}\label{T:lapp}
Let $a\ge 0$, $b< 0$, $c> 0$, and let $Y_t$ be a CIR process in Lemma \ref{L:sviaz1} such that $Y_0=y_0$ a.s. 
Then for every $\lambda> 0$,
\begin{align}
&\mathbb{E}_{y_0}\left[\exp\left\{-\lambda\int_0^tY_sds\right\}\right]  \nonumber \\
&=\exp\left\{-\frac{abt}{c^2}\right\}\left(\frac{\sqrt{b^2+2c^2\lambda}}
{\sqrt{b^2+2c^2\lambda}\cosh(\frac{1}{2}t\sqrt{b^2+2c^2\lambda})
-b\sinh(\frac{1}{2}t\sqrt{b^2+2c^2\lambda})}\right)^{\frac{2a}{c^2}} \nonumber \\
&\quad\exp\left\{-\frac{2y_0\lambda\sinh(\frac{1}{2}t\sqrt{b^2+2c^2\lambda})}
{\sqrt{b^2+2c^2\lambda}\cosh(\frac{1}{2}t\sqrt{b^2+2c^2\lambda})
-b\sinh(\frac{1}{2}t\sqrt{b^2+2c^2\lambda})}\right\}.
\label{E:lai2}
\end{align}
\end{theorem}

Theorem \ref{T:lapp} characterizes the Laplace transform of the probability distribution of the random variable $\int_0^tY_sds$. 
In the next section, we will study the asymptotic behavior of the inverse Laplace transform in a class of functions which look 
like the function on the right-hand side of (\ref{E:lai2}).
\section{An asymptotic inverse of the Laplace transform}\label{S:ilt}
Let us assume that 
$M$ is a function on $(0,\infty)$ whose Laplace transform is given by the following formula:
$$
\int_0^{\infty}e^{-\lambda y}M(y)dy=I(\lambda),\quad\lambda> 0,
$$
where 
$$
I(\lambda)=\lambda^{\gamma_1}G_1(\lambda)^{\gamma_2}G_2(\lambda)e^{F(\lambda)}
$$
with $\gamma_1\ge 0$ and $\gamma_2\ge 0$. We will next explain what restrictions are imposed on the functions 
$G_1$, $G_2$, and $F$. The function $G_2$ is analytic in the closed half-plane 
$\overline{\mathbb{C}}_{+}=\left\{\lambda:Re(\lambda)\ge 0\right\}$ and
such that $G_2(0)\neq 0$. It is also assumed that the function $G_1$ is analytic in $\overline{\mathbb{C}}_{+}$ except for a simple pole 
at $\lambda=0$ with residue $1$, that is, 
$G_1(\lambda)=\frac{1}{\lambda}+\widetilde{G}(\lambda)$
where $\widetilde{G}$ is an analytic function in $\overline{\mathbb{C}}_{+}$. In addition, we suppose that $G_1$ is a nowhere vanishing 
function in $\mathbb{C}_{+}=\left\{\lambda:Re(\lambda)> 0\right\}$. Similarly, the function $F$ is analytic in 
$\overline{\mathbb{C}}_{+}$ and has a simple pole at $\lambda=0$ with residue $\alpha> 0$, that is,
$F(\lambda)=\frac{\alpha}{\lambda}+\widetilde{F}(\lambda)$,
where $\widetilde{F}$ is an analytic function in $\overline{\mathbb{C}}_{+}$.
We also suppose that
\begin{equation}
\left|G_1(\lambda)^{\gamma_2}G_2(\lambda)e^{F(\lambda)}\right|\le\exp\left\{-|\lambda|^{\delta}\right\}
\label{E:lti2}
\end{equation}
as $|\lambda|\rightarrow\infty$ in $\overline{\mathbb{C}}_{+}$, for some $\delta> 0$ (much less is needed about the decay of the 
function in (\ref{E:lti2})).
\begin{remark}\label{E:stch}\rm
We assume that the function $G_1$ does not vanish in $\mathbb{C}_{+}$ in order to justify the existence of the power $G_1^{\gamma_2}$. 
Here we use the fact that for a nowhere vanishing analytic function $f$ in $\mathbb{C}_{+}$, there exists an analytic function $g$ on 
$\mathbb{C}_{+}$ such that $f(\lambda)=e^{g(\lambda)},\,\lambda\in\mathbb{C}_{+}$ (see Theorem 6.2 in \cite{keySS2}).
\end{remark}

The next result provides an asymptotic formula for the inverse Laplace transform of the function $I$.
\begin{theorem}\label{T:lti}
Suppose that the functions $G_1$, $G_2$, and $F$ are such as above. Then the following asymptotic formula holds:
\begin{equation}
M(y)=\frac{1}{2\sqrt{\pi}}\alpha^{\frac{1}{4}+\frac{\gamma_1-\gamma_2}{2}}G_2(0)e^{\widetilde{F}(0)}
y^{-\frac{3}{4}+\frac{\gamma_2-\gamma_1}{2}}e^{2\sqrt{\alpha}\sqrt{y}}\left(1+O\left(y^{-\frac{1}{4}}\right)\right)
\label{E:lti3}
\end{equation}
as $y\rightarrow\infty$.
\end{theorem}

\it Proof. \rm Using the Laplace transform inversion formula, we see that for every $\varepsilon> 0$ we have \\
$M(y)=\frac{1}{2\pi i}\int_{z=\varepsilon+ir}I(z)e^{yz}dz$.
It follows that
$M(\alpha y)=\frac{1}{2\pi i}\int_{z=\varepsilon+ir}I(z)e^{\alpha yz}dz$.
By (\ref{E:lti2}) and Cauchy's formula, we can deform the contour of integration into a new contour $\eta$ consisting of the following three parts: the half-line $(-\infty i,-y^{-\frac{1}{2}}i]$,
the half-circle $\Gamma$ in the right half-plane of radius $y^{-\frac{1}{2}}$ centered at $0$ (it is oriented counterclockwise), 
and finally the half-line $[y^{-\frac{1}{2}}i,\infty i)$. It follows that
\begin{align}
&M(\alpha y)=\frac{1}{2\pi i}\int_{\eta}I(z)e^{\alpha yz}dz 
=\frac{1}{2\pi}\int_{-\infty}^{-y^{-\frac{1}{2}}}
(ir)^{\gamma_1}G_1(ir)^{\gamma_2}G_2(ir)e^{\widetilde{F}(ir)}e^{-\frac{i\alpha}{r}}e^{i\alpha yr}dr \nonumber \\
&\quad+\frac{1}{2\pi i}\int_{\Gamma}z^{\gamma_1}G_1(z)^{\gamma_2}G_2(z)e^{\widetilde{F}(z)}e^{\frac{\alpha}{z}}
e^{\alpha yz}dz+\frac{1}{2\pi}\int_{y^{-\frac{1}{2}}}^{\infty}(ir)^{\gamma_1}G_1(ir)^{\gamma_2}G_2(ir)e^{\widetilde{F}(ir)}
e^{-\frac{i\alpha}{r}}e^{i\alpha yr}dr 
\nonumber \\
&=I_1(y)+I_2(y)+I_3(y).
\label{E:lti4}
\end{align}

We will first estimate $I_2(y)$. This will give the main contribution to the asymptotics. By making a substitution
$z=y^{-\frac{1}{2}}e^{i\theta}$, $-\frac{\pi}{2}\le\theta\le\frac{\pi}{2}$, we see that
\begin{align*}
I_2(y)&=\frac{1}{2\pi}y^{-\frac{1+\gamma_1}{2}}\int_{-\frac{\pi}{2}}^{\frac{\pi}{2}}
e^{i\theta\gamma_1}G_1\left(y^{-\frac{1}{2}}e^{i\theta}\right)^{\gamma_2}
G_2\left(y^{-\frac{1}{2}}e^{i\theta}\right) \\
&\quad\exp\left\{\widetilde{F}\left(y^{-\frac{1}{2}}e^{i\theta}\right)\right\}
\exp\left\{\alpha\sqrt{y}e^{-i\theta}\right\}
\exp\left\{\alpha\sqrt{y}e^{i\theta}\right\}e^{i\theta}d\theta.
\end{align*}
Next, taking into account the formula
$\sqrt{y}\left(e^{i\theta}+e^{-i\theta}\right)=2\sqrt{y}\cos\theta=2\sqrt{y}
+2\sqrt{y}(\cos\theta-1)$,
we obtain
\begin{align}
I_2(y)&=\frac{1}{2\pi}y^{-\frac{1+\gamma_1}{2}}e^{2\alpha\sqrt{y}}\int_{-\frac{\pi}{2}}^{\frac{\pi}{2}}
e^{i\theta\left(1+\gamma_1\right)}
G_1\left(y^{-\frac{1}{2}}e^{i\theta}\right)^{\gamma_2}G_2\left(y^{-\frac{1}{2}}e^{i\theta}\right) \nonumber \\
&\quad\exp\left\{\widetilde{F}\left(y^{-\frac{1}{2}}e^{i\theta}\right)\right\}
\exp\left\{2\alpha\sqrt{y}(\cos\theta-1)\right\}d\theta.
\label{E:I_2}
\end{align}
It is easy to see that
\begin{equation}
e^{i\theta\left(1+\gamma_1\right)}=1+O(\theta),
\label{E:thet}
\end{equation}
\begin{equation}
G_2\left(y^{-\frac{1}{2}}e^{i\theta}\right)-G_2(0)=O\left(y^{-\frac{1}{2}}\right),\quad
\exp\left\{\widetilde{F}\left(y^{-\frac{1}{2}}e^{i\theta}\right)\right\}-e^{\widetilde{F}(0)}
=O\left(y^{-\frac{1}{2}}\right)
\label{E:tlast2}
\end{equation}
on the contour $\Gamma$. Moreover, using (\ref{E:lti0}) and the mean value theorem, we obtain
\begin{align*}
G_1\left(y^{-\frac{1}{2}}e^{i\theta}\right)^{\gamma_2}-\left(\sqrt{y}e^{-i\theta}\right)^{\gamma_2}
=\left[\sqrt{y}e^{-i\theta}
+\widetilde{G}\left(y^{-\frac{1}{2}}e^{i\theta}\right)\right]^{\gamma_2}-\left(\sqrt{y}e^{-i\theta}\right)^{\gamma_2}
=O\left(y^{\frac{\gamma_2-1}{2}}\right)
\end{align*}
on $\Gamma$. Therefore, 
\begin{equation}
G_1\left(y^{-\frac{1}{2}}e^{i\theta}\right)^{\gamma_2}-y^{\frac{\gamma_2}{2}}=
O\left(y^{\frac{\gamma_2}{2}}|\theta|\right)+O\left(y^{\frac{\gamma_2-1}{2}}\right)
\label{E:gt}
\end{equation}
on $\Gamma$. It follows from (\ref{E:I_2}), (\ref{E:thet}), (\ref{E:tlast2}), and (\ref{E:gt}) that
\begin{align}
I_2(y)&=\frac{1}{2\pi}G_2(0)e^{\widetilde{F}(0)}
y^{-\frac{1+\gamma_1-\gamma_2}{2}}e^{2\alpha\sqrt{y}}\int_{-\frac{\pi}{2}}^{\frac{\pi}{2}}
\exp\left\{2\alpha\sqrt{y}(\cos\theta-1)\right\} \nonumber \\
&\quad\left(1+O\left(y^{-\frac{1}{2}}\right)+O(|\theta|)\right)d\theta.
\label{E:lappp}
\end{align}

We will next employ Laplace's method to estimate the integral appearing in (\ref{E:lappp}). 
Consider the integral $\displaystyle{\int_a^be^{-s\Phi(x)}\psi(x)dx}$, where 
$\Phi\in C^{\infty}[a,b]$ and $\psi\in C^{\infty}[a,b]$ (much less is needed from the functions $\Phi$ 
and $\psi$), and assume that there is an $x_0\in(a,b)$ 
such that $\Phi^{\prime}\left(x_0\right)=0$,
and $\Phi\left(x_0\right)> 0$ throughout $[a,b]$. Then the following assertion holds:
\begin{theorem}\label{T:Lapl}
Under the above assumptions, with $s> 0$ and $s\rightarrow\infty$,
\begin{equation}
\int_a^be^{-s\Phi(x)}\psi(x)dx=e^{-s\Phi(x_0)}\left[\frac{A}{\sqrt{s}}+O\left(\frac{1}{s}\right)\right],
\label{E:tlast5}
\end{equation}
where $A=\sqrt{2\pi}\psi\left(x_0\right)\left(\Phi^{\prime\prime}\left(x_0\right)\right)^{-\frac{1}{2}}$.
\end{theorem}
The proof of Theorem \ref{T:Lapl} can be found, e.g., in \cite{keySS2}.

Using (\ref{E:tlast5}) with $a=-\frac{\pi}{2}$, $b=\frac{\pi}{2}$, 
$\Phi(x)=1-\cos x$, $\psi(x)=1$, $x_0=0$, and $s=2\alpha\sqrt{y}$, we see that
\begin{equation}
\int_{-\frac{\pi}{2}}^{\frac{\pi}{2}}e^{2\alpha\sqrt{y}(\cos\theta-1)}d\theta=\frac{\sqrt{\pi}}{\sqrt{\alpha}}
y^{-\frac{1}{4}}+O\left(y^{-\frac{1}{2}}\right),\quad y\rightarrow\infty.
\label{E:tlast7}
\end{equation}
Similarly
\begin{equation}
\int_{-\frac{\pi}{2}}^{\frac{\pi}{2}}|\theta|e^{2\alpha\sqrt{y}(\cos\theta-1)}d\theta=O\left(y^{-\frac{1}{2}}\right),
\quad y\rightarrow\infty.
\label{E:tlast8}
\end{equation}
Therefore, (\ref{E:lappp}), (\ref{E:tlast7}), and (\ref{E:tlast8}) give
\begin{equation}
I_2(y)
=\frac{1}{2\sqrt{\pi\alpha}}G_2(0)e^{\widetilde{F}(0)}
y^{-\frac{3}{4}+\frac{\gamma_2-\gamma_1}{2}}e^{2\alpha\sqrt{y}}\left(1+O\left(y^{-\frac{1}{4}}\right)\right),
\quad y\rightarrow\infty.
\label{E:tlast9}
\end{equation}
Moreover, using (\ref{E:lti2}) we obtain
\begin{equation}
I_1(y)+I_3(y)=O\left(\exp\left\{-cy^{\delta}\right\}\right),\quad y\rightarrow\infty,
\label{E:tlast10}
\end{equation}
for some $c> 0$. It follows from (\ref{E:lti4}), (\ref{E:tlast9}), and (\ref{E:tlast10}) that
\begin{equation}
M\left(\alpha y\right)=\frac{1}{2\sqrt{\pi\alpha}}G_2(0)e^{\widetilde{F}(0)}
y^{-\frac{3}{4}+\frac{\gamma_2-\gamma_1}{2}}e^{2\alpha\sqrt{y}}\left(1+O\left(y^{-\frac{1}{4}}\right)\right),
\quad y\rightarrow\infty.
\label{E:UU}
\end{equation}
Now it is not hard to see that (\ref{E:UU}) implies (\ref{E:lti3}).

This completes the proof of Theorem \ref{T:lti}.
\section{Proof of Theorem \ref{T:oh}}\label{S:firh2} 
We will first discuss the properties of the following complex function:
\begin{equation}
\Phi_s(z)=z\,\cos z+s\,\sin z\quad,\quad z\in\mathbb{C},
\label{E:o}
\end{equation}
where $s\ge -1$. The next lemma concerns the zeros of $\Phi_s$. A special case of this lemma was stated in \cite{keySS1} without proof and was used in
 \cite{keyDY} and \cite{keySS1}. 
\begin{lemma}\label{L:seque}
For all $s\ge -1$, the function $\Phi_s$ has only real zeros.
\end{lemma}

\it Proof of Lemma \ref{L:seque}. \rm For every $n\ge 1$, put
$P_n(z)=z\prod_{k=1}^n\left(1-\frac{z^2}{k^2}\right)$.
The function $P_n$ is a polynomial of degree $2n+1$, all of whose roots ($z=k$, $k\in\mathbb{Z}$, $|k|\le n$) are real. Put
$Q_n(z)=z^{-s+1}\frac{d}{dz}\left(z^sP_n(z)\right)=sP_n(z)+zP_n^{\prime}(z)$.
Then $Q_n$ is a polynomial of degree $2n+1$ which vanishes at $z=0$. Also, by Rolle's theorem, the function 
$\displaystyle{\frac{d}{dz}\left(z^sP_n(s)\right)}$ vanishes at points strictly between $k$ and $k+1$, $-n\le k< n$,
since the function $z^sP_n(z)$ vanishes at those points. It follows from the previous considerations that $Q_n(z)$ has all its $2n+1$ roots that are real. But 
$\displaystyle{P_n(z)\rightarrow\frac{\sin\pi z}{\pi}}$ by the product formula. Hence,
$\displaystyle{Q_n(z)\rightarrow\frac{s}{\pi}\sin\pi z+z\cos\pi z}$,
and the desired conclusion that all the roots are real follows from the Rouch\'{e}-Hurwitz theorem.
The proof above implicitly used the condition $s> -1$ (for otherwise $z^sP_n(z)$ does not vanish at the origin).
The result for $s=-1$ can be derived from that for $s> -1$ by a limiting argument.

This completes the proof of Lemma \ref{L:seque}.

In the present paper, we consider only the case where $s\ge 0$. It is clear that the function $\Phi_s$ is odd and satisfies $\Phi_s(0)=0$. 
\begin{definition}\label{D:seq}
For $s\ge 0$, the smallest positive zero of the function $\Phi_s$ will be denoted by $r_s$.
\end{definition}

The number $r_s$ plays an important role throughout the paper. It is not hard to see that $r_0=\frac{\pi}{2}$, 
and $r_s\uparrow\pi$ as $s\rightarrow\infty$. 
Moreover, the function $s\mapsto r_s$ is differentiable and increasing on $(0,\infty)$. Indeed, the value of $r_s$ 
for $0< s<\infty$ is equal to the 
first coordinate of the point in $\mathbb{R}^2$ where the segment $y=-s^{-1}x$, $\frac{\pi}{2}< x<\pi$, intersects the curve 
$y=\tan x$. In addition, we have $r_s=\phi^{-1}(s)$, $0< s<\infty$, where $\phi(u)=-u(\tan u)^{-1}$, 
$\frac{\pi}{2}< u<\pi$. It is also clear that
$\sin(r_0)=1$, $\cos(r_0)=0$, and $\Phi^{\prime}_0\left(r_0\right)=-\frac{\pi}{2}$.
Moreover, if $s> 0$, then 
\begin{equation}
\sin(r_s)> 0,\quad\cos(r_s)< 0,\quad\mbox{and}\quad\Phi^{\prime}_s\left(r_s\right)< 0.
\label{E:urr2}
\end{equation}
By Lemma \ref{L:seque}, the function $\rho_s$ defined by
\begin{equation}
\rho_s(z)=z\cosh z+s\sinh z=-i\Phi_s(iz)
\label{E:fro}
\end{equation}
has only imaginary zeros. 

The next lemma concerns the mixing distribution densities.
\begin{lemma}\label{L:disden}
Suppose that $a\ge 0$, $b< 0$, and $c>0$ in the Heston model. Then the probability distribution of the random variable 
$\alpha_t$ in (\ref{E:mix2}) admits a density $m_t$ for every $t> 0$.
\end{lemma}

\it Proof. \rm 
We will first prove that the probability distribution of the random variable $\int_0^tY_sds$ 
admits a density $\overline{m}_t$. This will allow us to prove the existence of the mixing distribution $m_t$, 
since $m_t$ can be determined from the formula
\begin{equation}
\overline{m}_t(y)=\frac{1}{2\sqrt{ty}}m_t\left(t^{-\frac{1}{2}}y^{\frac{1}{2}}\right).
\label{E:hof}
\end{equation}

It follows from Theorem \ref{T:lapp} that
\begin{align}
&\mathbb{E}_{y_0}\left[\exp\left\{-\frac{\lambda}{2c^2}\int_0^tY_sds\right\}\right]
=\exp\left\{-\frac{abt}{c^2}\right\}\left(\frac{\sqrt{b^2+\lambda}}
{\sqrt{b^2+\lambda}\cosh(\frac{1}{2}t\sqrt{b^2+\lambda})
-b\sinh(\frac{1}{2}t\sqrt{b^2+\lambda})}\right)^{\frac{2a}{c^2}} \nonumber \\
&\quad\exp\left\{-\frac{y_0c^{-2}\lambda\sinh(\frac{1}{2}t\sqrt{b^2+\lambda})}
{\sqrt{b^2+\lambda}\cosh(\frac{1}{2}t\sqrt{b^2+\lambda})
-b\sinh(\frac{1}{2}t\sqrt{b^2+\lambda})}\right\}.
\label{E:laika}
\end{align}
Denote by $\Psi_1$ and $\Psi_2$ the functions on the right-hand side and the left-hand side of formula (\ref{E:laika}), respectively,
and put
\begin{equation}
u_{b,t}=-4t^{-2}r_{\frac{1}{2}t|b|}^2
\label{E:ver}
\end{equation} 
where $t>0$ and $b\le 0$. It is not hard to see that the function $\Psi_1$ can be continued analytically from the half-line to the  
half-plane 
\begin{equation}
\mathbb{C}_{b,t}=\left\{\lambda\in\mathbb{C}:\,Re(\lambda)>-b^2+u_{b,t}\right\}.
\label{E:fof}
\end{equation}
This can be shown using the properties of the zeros of the function $\rho_s$ defined by (\ref{E:fro}) and the fact that 
$\lambda=0$ is not a singularity of the function $\Psi_1$. 
On the other hand, the function $\Psi_2$ is analytic in the half-plane 
$\mathbb{C}_{+}=\left\{\lambda\in\mathbb{C}:\,Re(\lambda)> 0\right\}$ (use formula 
(\ref{E:laika}) with $\lambda> 0$ to prove the finiteness of the derivative of the function $\Psi_2$ in $\mathbb{C}_{+}$). It follows that 
formula (\ref{E:laika}) holds for all $\lambda\in\mathbb{C}_{b,t}$. 

Since $i\xi\in \mathbb{C}_{b,t}$ for all $\xi\in\mathbb{R}^1$, equation (\ref{E:laika}) implies the following formula:
\begin{equation}
\int_0^{\infty}e^{-i\xi x}d\nu_t(x)=\Psi_1(i\xi),\quad\xi\in\mathbb{R}^1,
\label{E:vera}
\end{equation}
where $\nu_t$ stands for the probability distribution of the random variable $\frac{1}{c^2}\int_0^tY_sds$. 

It is not hard to see that
$$
\Psi_1(\lambda)-c_1\left(\frac{z}{\rho_s(z)}\right)^{\frac{2a}{c^2}}\exp\left\{-\frac{\left(c_2z^2-c_3\right)\sinh z}
{\rho_s(z)}\right\}
$$
where $z=\frac{1}{2}t\sqrt{b^2+\lambda}$, $s=-b$, $\rho_s$ is defined by (\ref{E:fro}), and $c_1$, $c_2$, and $c_3$ are some positive constants.

Next, we observe that for $z$ lying in a proper sector of the right-hand plane and $z\rightarrow\infty$, we have 
$$
\left(c_2z^2-c_3\right)\frac{\sinh z}{\rho_s(z)}=c_2z+O(1)
$$
and
$$
\frac{z}{\rho_s(z)}=2e^{-z}+O\left(|e^{-2z}|\right).
$$
This implies that
\begin{equation}
\left|\Psi_1(\lambda)\right|\le C_1\exp\left\{-C_2|\lambda|^{\frac{1}{2}}\right\}
\label{E:tut}
\end{equation}
for $\lambda\in\mathbb{C}_{b,t}$ and $\lambda\rightarrow\infty$.

It follows from (\ref{E:tut})  that the function $\xi\mapsto\left|\Psi_1(i\xi)\right|$ belongs to
the space $L^2\left(\mathbb{R}^1\right)$. Taking into account (\ref{E:vera}) and using Plancherel's Theorem, we see that the measure
$\nu_t$ is absolutely continuous with respect to the Lebesgue measure on $\mathbb{R}^1$. Therefore, the density $\bar{m}_t$ exists. This implies the 
existence of the mixing density $m_t$, and the proof of Lemma \ref{L:disden} is thus completed.

We will next prove Theorem \ref{T:oh}. The following lemma will be used in the proof. In this lemma, we use the notation 
introduced in (\ref{E:o}) and (\ref{E:ver}).
\begin{lemma}\label{L:hers}
Let $a\ge 0$, $b< 0$, $c> 0$, and let $Y_t$ be a CIR process satisfying the equation \\
$dY_t=\left(a+bY_t\right)dt+c\sqrt{Y_t}dZ_t$ and such that $Y_0=y_0$ a.s. 
Then the following formula holds:
\begin{align}
&\int_0^{\infty}e^{-\lambda y}y^{-\frac{1}{2}}\exp\left\{\left(b^2-u_{b,t}\right)y\right\}
m_t\left(\frac{\sqrt{2}c}{\sqrt{t}}\sqrt{y}\right)dy 
=\frac{\sqrt{2t}}{c}\exp\left\{-\frac{abt}{c^2}\right\}
\left(\frac{it\sqrt{\lambda+u_{b,t}}}{2\Phi_{\frac{1}{2}t|b|}\left(i\frac{1}{2}t\sqrt{\lambda+u_{b,t}}\right)}\right)^{\frac{2a}{c^2}} \nonumber \\
&\quad
\exp\left\{-\frac{ity_0\left(\lambda+u_{b,t}-b^2\right)\sinh\left(\frac{1}{2}t\sqrt{\lambda+u_{b,t}}\right)}
{2c^2\Phi_{\frac{1}{2}t|b|}\left(i\frac{1}{2}t\sqrt{\lambda+u_{b,t}}\right)}\right\},
\label{E:laai}
\end{align}
The functions on the both sides of (\ref{E:laai}) are analytic in the right half-plane 
$\mathbb{C}_0=\left\{\lambda\in\mathbb{C}:\,Re(\lambda)>0\right\}$, and the equality in (\ref{E:laai}) holds for all
$\lambda\in\mathbb{C}_0$.
\end{lemma}

\it Proof. \rm
It was established above that formula (\ref{E:laika}) holds for all $\lambda\in\mathbb{C}_{b,t}$, where $\mathbb{C}_{b,t}$ 
is the half-plane defined by (\ref{E:fof}). It is not hard to see that Lemma \ref{L:hers} follows from (\ref{E:laika}), 
Lemma \ref{L:disden}, and (\ref{E:hof}). 

We will next compute the residue $\lambda_0$ of the function 
\begin{align}
\Lambda(\lambda)&=\frac{\sqrt{\lambda+u_{b,t}}}{\sqrt{\lambda+u_{b,t}}\cosh\left(\frac{1}{2}t\sqrt{\lambda+u_{b,t}}\right)
+|b|\sinh\left(\frac{1}{2}t\sqrt{\lambda+u_{b,t}}\right)} \nonumber \\    
&=\frac{it\sqrt{\lambda+u_{b,t}}}{2\Phi_{\frac{1}{2}t|b|}\left(i\frac{1}{2}t\sqrt{\lambda+u_{b,t}}\right)}
\label{E:lll}
\end{align}
at $\lambda=0$. Since 
\begin{equation}
\Phi^{\prime}_{\frac{1}{2}t|b|}\left(r_{\frac{1}{2}t|b|}\right)=\left(1+\frac{1}{2}t|b|\right)\cos\left(r_{\frac{1}{2}t|b|}\right)
-r_{\frac{1}{2}t|b|}\sin\left(r_{\frac{1}{2}t|b|}\right),
\label{E:fofa}
\end{equation}
it is not hard to see that
\begin{align}
&\lambda_0=\lim_{\lambda\rightarrow 0}\lambda\Lambda(\lambda)=-r_{\frac{1}{2}t|b|}\lim_{\lambda\rightarrow 0}
\frac{\lambda}{\Phi_{\frac{1}{2}t|b|}\left(i\frac{1}{2}t\sqrt{\lambda+u_{b,t}}\right)} \nonumber \\
&=-8t^{-2}r^2_{\frac{1}{2}t|b|}\Phi^{\prime}_{\frac{1}{2}t|b|}\left(r_{\frac{1}{2}t|b|}\right)^{-1}
=8t^{-2}r^2_{\frac{1}{2}t|b|}\left|\Phi^{\prime}_{\frac{1}{2}t|b|}\left(r_{\frac{1}{2}t|b|}\right)\right|^{-1} \nonumber \\
&=8t^{-2}r^2_{\frac{1}{2}t|b|}\left|\left(1+\frac{1}{2}t|b|\right)\cos\left(r_{\frac{1}{2}t|b|}\right)
-r_{\frac{1}{2}t|b|}\sin\left(r_{\frac{1}{2}t|b|}\right)\right|^{-1}.
\label{E:qqq}
\end{align}
In the proof of (\ref{E:qqq}), we used (\ref{E:ver}) and (\ref{E:urr2}).

Our next goal is to apply Theorem \ref{T:lti} to the Laplace transform in formula (\ref{E:laai}). The numbers $\gamma_1$ and 
$\gamma_2$ and the functions $G_1$, $G_2$, and $F$ in Theorem \ref{T:lti} are chosen as follows: 
$\gamma_1=0$, $\gamma_2=\frac{2a}{c^2}$, and $G_1(\lambda)=\frac{1}{\lambda_0}\Lambda(\lambda)$,
where $\Lambda$ and $\lambda_0$ are defined by (\ref{E:lll}) and (\ref{E:qqq}), 
$G_2(\lambda)=\lambda_0^{\frac{2a}{c^2}}\frac{\sqrt{2t}}{c}\exp\left\{-\frac{abt}{c^2}\right\}$,
and
$$
F(\lambda)=-\frac{ity_0\left(\lambda+u_{b,t}-b^2\right)\sinh\left(\frac{1}{2}t\sqrt{\lambda+u_{b,t}}\right)}
{2c^2\Phi_{\frac{1}{2}t|b|}\left(i\frac{1}{2}t\sqrt{\lambda+u_{b,t}}\right)}.
$$
Note that $G_1$ is a nowhere vanishing function in $\mathbb{C}_{+}$.It follows from (\ref{E:tut}) that the decay condition in 
(\ref{E:lti2}) is satisfied. The next lemma provides explicit formulas for the residue 
$\alpha$ of the function $F$ at $\lambda=0$ and for the number $\widetilde{F}(0)$. 
\begin{lemma}\label{L:alfo}
The following formulas hold:
\begin{equation}
\alpha=\frac{ty_0\eta_1}{2c^2|\rho_1|}\left(b^2+4t^{-2}r_{\frac{t|b|}{2}}^2\right)
> 0.
\label{E:lti0}
\end{equation}
and
\begin{equation}
\widetilde{F}(0)=\frac{ty_0}{2c^2\rho_1^2}\left[\left(\eta_1-\left(4t^{-2}r^2_{\frac{1}{2}t|b|}+b^2\right)\eta_2\right)\rho_1
+\left(4t^{-2}r^2_{\frac{1}{2}t|b|}+b^2\right)\rho_2\right].
\label{E:lti00}
\end{equation}
The constants $\eta_1$, $\eta_2$, $\rho_1$, and $\rho_2$  in (\ref{E:lti0})
and (\ref{E:lti00}) are given by
\begin{equation}
\eta_1=\sin\left(r_{\frac{1}{2}t|b|}\right),\quad\eta_2=-\frac{t^2\cos\left(r_{\frac{1}{2}t|b|}\right)}{8r_{\frac{1}{2}t|b|}},
\label{E:eta}
\end{equation}
\begin{equation}
\rho_1=\frac{t^2}{8r_{\frac{1}{2}t|b|}}\left[\left(1+\frac{1}{2}t|b|\right)\cos\left(r_{\frac{1}{2}t|b|}\right)
-r_{\frac{1}{2}t|b|}\sin\left(r_{\frac{1}{2}t|b|}\right)\right],
\label{E:rho1}
\end{equation}
and
\begin{equation}
\rho_2=\frac{t^4}{128}\left[r^{-3}_{\frac{1}{2}t|b|}\left[\left(1+\frac{1}{2}t|b|\right)\cos\left(r_{\frac{1}{2}t|b|}\right)
-r_{\frac{1}{2}t|b|}\sin\left(r_{\frac{1}{2}t|b|}\right)\right]+2r^{-2}_{\frac{1}{2}t|b|}\sin\left(r_{\frac{1}{2}t|b|}
\right)\right].
\label{E:rho2}
\end{equation}
\end{lemma}

\it Proof. \rm We will need the first two coefficients in the power series representation for the function $\lambda F(\lambda)$.
It is not hard to see that 
\begin{equation}
\sinh\left(\frac{1}{2}t\sqrt{\lambda+u_{b,t}}\right)=i\eta_1+i\eta_2\lambda+\cdots
\label{E:bbi2}
\end{equation}
where $\eta_1$ and $\eta_2$ are defined by (\ref{E:eta}).
Next, using formula (\ref{E:fofa}) and taking into account that
$$
\Phi^{\prime\prime}_{\frac{1}{2}t|b|}\left(r_{\frac{1}{2}t|b|}\right)=-\Phi_{\frac{1}{2}t|b|}\left(r_{\frac{1}{2}t|b|}\right)
-2\sin\left(r_{\frac{1}{2}t|b|}\right)=-2\sin\left(r_{\frac{1}{2}t|b|}\right),
$$
we obtain
\begin{equation}
\Phi_{\frac{1}{2}t|b|}\left(i\frac{1}{2}t\sqrt{\lambda+u_{b,t}}\right)
=\rho_1\lambda+\rho_2\lambda^2+\cdots
\label{E:bbi1}
\end{equation}
where $\rho_1$ and $\rho_2$ are defined by (\ref{E:rho1}) and (\ref{E:rho2}). 
Using (\ref{E:urr2}), we see that $\eta_1> 0$ and $\eta_2> 0$. Moreover,
(\ref{E:bbi2}), (\ref{E:bbi1}), and the definition of $F$ imply the
following equality:
\begin{align}
\lambda
F(\lambda)&=\frac{ty_0}{2c^2}\frac{\left(-4t^{-2}r^2_{\frac{1}{2}t|b|}-b^2+\lambda\right)
\left(\eta_1+\eta_2
\lambda+\cdots\right)}{\rho_1+\rho_2\lambda+\cdots} \nonumber \\
&=\frac{ty_0}{2c^2}\frac{\left(-4t^{-2}r^2_{\frac{1}{2}t|b|}-b^2\right)\eta_1
+\left(\eta_1-\left(4t^{-2}r^2_{\frac{1}{2}t|b|}+b^2\right)\eta_2\right)\lambda+\cdots}
{\rho_1+\rho_2\lambda+\cdots}
\label{E:sco}
\end{align}
where $\rho_1$, $\rho_2$, $\eta_1$, and $\eta_2$ are defined in (\ref{E:rho1}), (\ref{E:rho2}),
and (\ref{E:eta}).

Now, it is clear that Lemma \ref{L:alfo} follows from (\ref{E:sco}).

We will next complete the proof of Theorem \ref{T:oh} and compute the constants appearing in it.
By applying Theorem \ref{T:lti} to the Laplace transform in (\ref{E:laai}), we see that
\begin{align}
&y^{-\frac{1}{2}}\exp\left\{\left(b^2-u_{b,t}\right)y\right\}
m_t\left(\frac{\sqrt{2}c}{\sqrt{t}}\sqrt{y};a,b,c,y_0\right) \nonumber \\
&=\frac{1}{2\sqrt{\pi}}\alpha^{\frac{1}{4}-\frac{a}{c^2}}
\lambda_0^{\frac{2a}{c^2}}\frac{\sqrt{2t}}{c}\exp\left\{-\frac{abt}{c^2}\right\}e^{\widetilde{F}(0)}y^{-\frac{3}{4}+\frac{a}{c^2}}
e^{2\sqrt{\alpha}\sqrt{y}}\left(1+O\left(y^{-\frac{1}{4}}\right)\right)
\label{E:obshch2}
\end{align}
as $y\rightarrow\infty$. Next, replacing $y$ by $y^2\frac{t}{2c^2}$ in formula (\ref{E:obshch2}), we obtain
\begin{align}
m_t\left(y;a,b,c,y_0\right)&=\frac{1}{2\sqrt{\pi}}\alpha^{\frac{1}{4}-\frac{a}{c^2}}
\lambda_0^{\frac{2a}{c^2}}\frac{\sqrt{2t}}{c}\exp\left\{-\frac{abt}{c^2}\right\}e^{\widetilde{F}(0)}
\left(\frac{t}{2c^2}\right)^{-\frac{1}{4}+\frac{a}{c^2}} \nonumber \\
&\quad y^{-\frac{1}{2}+\frac{2a}{c^2}}
\exp\left\{\frac{\sqrt{2\alpha t}}{c}y\right\}\exp\left\{-\frac{t\left(b^2-u_{b,t}\right)}{2c^2}y^2\right\}
\left(1+O\left(y^{-\frac{1}{2}}\right)\right)
\label{E:obshch22}
\end{align}
as $y\rightarrow\infty$.  Now it is clear that (\ref{E:obshch22}) implies Theorem \ref{T:oh}. In addition, 
the following lemma describes the constants appearing in Theorem \ref{T:oh}:
\begin{lemma}\label{L:con1}
The constants $A$, $B$, and $C$ in Theorem \ref{T:oh} are given by 
$$
A=\frac{1}{\sqrt{\pi}}\left(\frac{t}{2c^2}\right)^{\frac{1}{4}+\frac{a}{c^2}}
\alpha^{\frac{1}{4}-\frac{a}{c^2}}\lambda_0^{\frac{2a}{c^2}}\exp\left\{-\frac{abt}{c^2}\right\}e^{\widetilde{F}(0)},
$$
$$
B=c^{-1}\sqrt{2\alpha t},\quad\mbox{and}\quad
C=\left(2c^2\right)^{-1}t\left(b^2+4t^{-2}r_{\frac{t|b|}{2}}^2\right),
$$
where the numbers $\lambda_0> 0$, $\alpha> 0$, and $\widetilde{F}(0)$ are defined 
in (\ref{E:qqq}), (\ref{E:lti0}), and (\ref{E:lti00}), respectively.
\end{lemma}

This completes the proof of Theorem \ref{T:oh}.
\begin{corollary}\label{C:Hes}
The following formula holds:
$$
m_t\left(y,a,0,c,y_0\right)=Ae^{-Cy^2}e^{By}y^{-\frac{1}{2}+\frac{2a}{c^2}}
\left(1+O\left(y^{-\frac{1}{2}}\right)\right)
$$
as $y\rightarrow\infty$, where
$A=\frac{2^{\frac{1}{4}+\frac{a}{c^2}}y_0^{\frac{1}{4}-\frac{a}{c^2}}}{c\sqrt{t}}\exp\left\{\frac{4y_0}{c^2t}\right\}$,
$B=\frac{2\sqrt{2y_0}\pi}{c^2t}$, and $C=\frac{\pi^2}{2c^2t}$.
\end{corollary}

Corollary \ref{C:Hes} can be derived from Theorem \ref{T:oh} and Lemma \ref{L:con1} by using the equality $r_0=\frac{\pi}{2}$
and the fact that for 
$b=0$ we have 
$\displaystyle{\lambda_0=\frac{4\pi}{t^2}}$, $\displaystyle{\alpha=\frac{4y_0\pi^2}{c^2t^3}}$, and 
$\displaystyle{\widetilde{F}(0)=-\frac{3y_0}{c^2t}}$.
\section{Proof of Theorem \ref{T:main}}\label{S:proofs}
We will next formulate a theorem concerning the asymptotic behavior of certain integral operators. This result 
is a minor modification of Theorem 4.5 established in \cite{keyGS2}. 
\begin{theorem}\label{T:into}
Let $A$, $\zeta$, and $b$ be positive Borel functions on $[0,\infty)$, and 
suppose that the following conditions hold:
\begin{enumerate}
\item The function $A$ is integrable over any finite sub-interval of $[0,\infty)$. 
\item The function $b$ is bounded and $\displaystyle{\lim_{y\rightarrow\infty}b(y)=0}$. 
\item There exist $y_1> 0$, $c> 0$, and $\gamma$ with $0<\gamma\le 1$ such that $\zeta$ and $b$ are differentiable 
on $[y_1,\infty)$, and in addition $\left|\zeta^{\prime}(y)\right|\le cy^{-\gamma}\zeta(y)$ and $\left|b^{\prime}(y)\right|\le cy^{-\gamma}b(y)$ 
for all $y\ge y_1$.
\item For every $a> 0$, there exists $y_a> 0$ such that $b(y)\zeta(y)\ge\exp\left\{-ay^4\right\}$, $y> y_a$.
\item There exists a real number $l$ such that $A(y)=e^{ly}\zeta(y)(1+O(b(y)))$ as $y\rightarrow\infty$. 
\end{enumerate}
Then, for every fixed $k> 0$, 
\begin{align*}
&\int_0^{\infty}A(y)\exp\left\{-\left(\frac{w^2}{y^2}+k^2y^2\right)\right\}dy  \\
&=\frac{\sqrt{\pi}}{2k}
\exp\left\{\frac{l^2}{16k^2}\right\}
\zeta\left(k^{-\frac{1}{2}}w^{\frac{1}{2}}\right)\exp\left\{lk^{-\frac{1}{2}}w^{\frac{1}{2}}\right\}
e^{-2kw}\left[1+O\left(w^{-\frac{\gamma}{2}}\right)+O\left(b\left(k^{-\frac{1}{2}}w^{\frac{1}{2}}\right)\right)\right],
\quad w\rightarrow\infty.
\end{align*}
\end{theorem}

The only difference between Theorem \ref{T:into} formulated above and Theorem 4.5 in \cite{keyGS2} is that in Theorem \ref{T:into} 
we do not assume the integrability of the function $\zeta$ near zero. It is not hard to see that Theorem \ref{T:into} 
can be derived from Theorem 4.5 if we replace the function $\zeta(y)$ by the function $A(y)e^{-ly}$ 
near zero. 

Let $A(y)=y^{-1}m_t(y)e^{Cy^2}$, $k=\sqrt{C+\frac{t}{8}}$, $l=B$, $\zeta(y)=Ay^{-\frac{3}{2}+\frac{2a}{c^2}}$,
and $b(y)=y^{-\frac{1}{2}}$, where the constants $A$, $B$, and $C$ are as in Theorem \ref{T:oh}.
Then, it is not hard to see that condition 2 in Theorem \ref{T:into} follows from formula (\ref{E:oh1}). 
In addition, it is clear that condition 3 with $\gamma=1$ holds. The next lemma 
shows that condition 1 in Theorem \ref{T:into} also holds.
\begin{lemma}\label{L:dodo}
For every $s> 0$, $\displaystyle{\int_0^sy^{-1}m_t(y)dy<\infty}$.
\end{lemma}

\it Proof. \rm The function on the right-hand side of 
(\ref{E:laai}) is integrable with respect to $\lambda$ over the interval $(1,\infty)$. 
Suppose $h$ is any positive function on $[0,\infty)$ such that its Laplace transform has this property. 
Then $\int_0^sh(y)y^{-1}dy<\infty$ for all $s> 0$.
It follows from this fact and (\ref{E:laai}) that the function 
$y\mapsto y^{-\frac{3}{2}}m_t\left(\frac{\sqrt{2}c}{\sqrt{t}}\sqrt{y}\right)$ 
is integrable over any interval of the form $[0,s]$ with $s> 0$. This implies 
Lemma \ref{L:dodo}. 

Next, applying Theorem \ref{T:into} we see that
\begin{align}
&\int_0^{\infty}y^{-1}m_t(y)\exp\left\{-\left(\frac{z^2}{y^2}+\frac{ty^2}{8}\right)\right\}dy \nonumber \\
&=A\frac{\sqrt{\pi}}{2k}\exp\left\{\frac{B^2}{16k^2}\right\}k^{\frac{3}{4}-\frac{a}{c^2}}z^{-\frac{3}{4}+\frac{a}{c^2}}
\exp\left\{Bk^{-\frac{1}{2}}\sqrt{z}\right\}e^{-2kz}\left(1+O\left(z^{-\frac{1}{4}}\right)\right)
\label{E:dop2}
\end{align}
as $z\rightarrow\infty$. Replacing $z$ by $\frac{\log x}{\sqrt{2t}}$ in 
formula (\ref{E:dop2}) and taking into account formula (\ref{E:stvoll1}) and the equality 
$\displaystyle{k=\frac{\sqrt{8C+t}}{2\sqrt{2}}}$, we obtain
\begin{align}
&D_t\left(x_0e^{\mu t}x\right)=\frac{A}{x_0e^{\mu t}}2^{-\frac{3}{4}+\frac{a}{c^2}}t^{-\frac{1}{8}-\frac{a}{2c^2}}
(8C+t)^{-\frac{1}{8}-\frac{a}{2c^2}}\exp\left\{\frac{B^2}{2(8C+t)}\right\}
 \nonumber \\
&\quad(\log x)^{-\frac{3}{4}+\frac{a}{c^2}}
\exp\left\{\frac{B\sqrt{2}}{t^{\frac{1}{4}}(8C+t)^{\frac{1}{4}}}\sqrt{\log x}\right\}x^{-\left(\frac{3}{2}+
\frac{\sqrt{8C+t}}{2\sqrt{t}}\right)}
\left(1+O\left((\log x)^{-\frac{1}{4}}\right)\right)
\label{E:dop3}
\end{align}
as $x\rightarrow\infty$.

Now it is clear that formula (\ref{E:dop3}) implies Theorem \ref{T:main}. Moreover, the following lemma holds:
\begin{lemma}\label{L:con2}
The constants $A_1$, $A_2$, and $A_3$ in Theorem \ref{T:main} are given by
$$
A_1=\frac{A}{x_0e^{\mu t}}2^{-\frac{3}{4}+\frac{a}{c^2}}t^{-\frac{1}{8}-\frac{a}{2c^2}}
(8C+t)^{-\frac{1}{8}-\frac{a}{2c^2}}\exp\left\{\frac{B^2}{2(8C+t)}\right\},
$$
$A_2=\frac{B\sqrt{2}}{t^{\frac{1}{4}}(8C+t)^{\frac{1}{4}}}$, and $A_3=\frac{3}{2}+
\frac{\sqrt{8C+t}}{2\sqrt{t}}$,
where $A$, $B$, and $C$ are defined in Lemma \ref{L:con1}.
\end{lemma}

Let $p\in\mathbb{R}$, and denote by $l_p$ the moment of order $p$ of the function $D_t$, that is, \\
$l_p=\mathbb{E}\left[X_t^p\right]=\int_0^{\infty}x^pD_t(x)dx$.
The next result was obtained in \cite{keyAP}. 
\begin{lemma}\label{L:blup}
For $a\ge 0$, $b\le 0$, $c> 0$, and $p\in\mathbb{R}$, the following statement holds: $q_p<\infty$ if and only if
$2-A_3< p< A_3-1$,
where the constant $A_3$ is such as in Theorem \ref{T:main}. For $b=0$, $q_p<\infty$ if and only if
$$
\frac{1}{2}-\frac{\sqrt{4\pi^2+c^2t^2}}{2ct}< p<\frac{1}{2}+\frac{\sqrt{4\pi^2+c^2t^2}}{2ct}.
$$
\end{lemma}
It is not hard to see that Lemma \ref{L:blup} and more precise integrability theorems for the distribution of the stock price follow 
from Theorem \ref{T:main}. 
\section{Proof of Theorem \ref{T:ohh}}\label{S:progen}
Let $Y_t$ be the volatility process in model (\ref{E:SS}). Then $Y_t^2$ is a squared Ornstein-Uhlenbeck process. The Laplace transform of the 
law of the squared Ornstein-Uhlenbeck process was found by Wenocur \cite{keyW} and by Stein and Stein \cite{keySS1}. 
Another explicit expression for this Laplace transform is given in the next formula:
\begin{align}
&\mathbb{E}_{y_0}\left[\exp\left\{-\lambda\int_0^tY_s^2ds\right\}\right]
=2\sqrt{t}e^{\frac{qt}{2}}\left(\frac{\sqrt{w}}{\sqrt{w}\cosh\left(t\sqrt{w}\right)+q\sinh\left(t\sqrt{w}\right)}
\right)^{\frac{1}{2}} \nonumber \\
&\quad\exp\left\{-\frac{y_0^2\lambda\sinh\left(t\sqrt{w}\right)}{\sqrt{w}\cosh\left(t\sqrt{w}\right)
+q\sinh\left(t\sqrt{w}\right)}\right\}
\exp\left\{-\frac{2mqy_0\lambda\left(\cosh\left(t\sqrt{w}\right)-1\right)}{\sqrt{w}
\left(\sqrt{w}\cosh\left(t\sqrt{w}\right)+q
\sinh\left(t\sqrt{w}\right)\right)}\right\} \nonumber \\
&\quad\exp\left\{\frac{m^2q^2\lambda\left(\sinh\left(t\sqrt{w}\right)-t\sqrt{w}\cosh\left(t\sqrt{w}\right)\right)}
{w\left(\sqrt{w}\cosh\left(t\sqrt{w}\right)+q
\sinh\left(t\sqrt{w}\right)\right)}\right\} \nonumber \\
&\quad\exp\left\{\frac{m^2q^3\lambda
\left(4\sinh^2\left(\frac{t\sqrt{w}}{2}\right)-t\sqrt{w}
\sinh\left(t\sqrt{w}\right)\right)}{w^{\frac{3}{2}}\left(\sqrt{w}\cosh\left(t\sqrt{w}\right)+q
\sinh\left(t\sqrt{w}\right)\right)}\right\}
\label{E:ochdli}
\end{align}
where $\lambda> 0$ and $w=q^2+2\sigma^2\lambda$. We will next sketch the proof of formula (\ref{E:ochdli}). 
The first step is to replace the symbols $\delta$, $\theta$, 
$k$, $\sigma_0$, and $\lambda$ in formula (8) in \cite{keySS1} by $q$, $m$, $\sigma$, $y_0$, and $t\lambda$, respectively, and take 
into account the following relations between the notation in \cite{keySS1} and in the present paper: $A=-\frac{q}{\sigma^2}$, 
$B=\frac{mq}{\sigma^2}$, $C=-\frac{\lambda}{\sigma^2}$, $a=\frac{1}{\sigma^2}\sqrt{q^2+2\sigma^2\lambda}$, 
$b=\frac{q}{\sqrt{q^2+2\sigma^2\lambda}}$, and $ak^2t=t\sqrt{q^2+2\sigma^2\lambda}$. We also combine the terms 
$\frac{a-A}{2a^2}a^2k^2t$ and $-\frac{1}{2}\log\left\{\frac{1}{2}\left(\frac{A}{a}+1\right)+\frac{1}{2}
\left(1-\frac{A}{a}\right)e^{2ak^2t}\right\}$
in the expression for $N$ in formula (7) in \cite{keySS1}, and after somewhat long and tedious computations show that formula 
(8) in \cite{keySS1} and formula (\ref{E:ochdli}) in the present paper are equivalent. 
\begin{remark}\label{R:mixis}\rm Lemma \ref{L:disden} states that for the Heston model, there exists the mixing distribution density $m_t$ 
for every $t> 0$. The same statement holds for the Stein-Stein model. 
This can be established using formula (\ref{E:ochdli}) and reasoning as in the proof of Lemma \ref{L:disden}.
\end{remark}

It follows from Remark \ref{R:mixis} that 
\begin{equation}
\mathbb{E}_{y_0}\left[\exp\left\{-\lambda\int_0^tY_s^2ds\right\}\right]=
\int_0^{\infty}e^{-\lambda ty^2}m_t(y)dy=\frac{1}{2\sqrt{t}}
\int_0^{\infty}e^{-\lambda y}y^{-\frac{1}{2}}m_t\left(t^{-\frac{1}{2}}y^{\frac{1}{2}}\right)dy.
\label{E:otchi}
\end{equation}
We will next obtain a sharp asymptotic formula for the mixing distribution density in the Stein-Stein model using formulas (\ref{E:ochdli}), 
(\ref{E:otchi}), and the methods developed in Section \ref{S:firh2}.

Recall that for $s\ge 0$, we denoted by $r_s$ the smallest strictly positive zero of the function 
$\Phi_s(z)=z\cos z+s\sin z$. It is clear that $z\cosh z+s\sinh z=-i\Phi_s(iz)$. For $q> 0$ and $t> 0$, 
put $v_{q,t}=-\frac{r^2_{qt}}{t^2}$. Now (\ref{E:ochdli}) and (\ref{E:otchi}) give 
\begin{align}
&\int_0^{\infty}e^{-\lambda y}y^{-\frac{1}{2}}\exp\left\{\left(q^2-v_{q,t}\right)y\right\}
m_t\left(\frac{\sqrt{2}\sigma\sqrt{y}}{\sqrt{t}}\right)dy \nonumber \\
&=\frac{\sqrt{2t}}{\sigma}\exp\left\{\frac{qt}{2}\right\}
F_1(\lambda)\exp\left\{F_2(\lambda)+F_3(\lambda)+F_4(\lambda)+F_5(\lambda)\right\},
\label{E:expo1}
\end{align}
where
\begin{equation}
F_1(\lambda)=\left(\frac{it\sqrt{\lambda+v_{q,t}}}{\Phi_{qt}\left(it\sqrt{\lambda+v_{q,t}}\right)}\right)^{\frac{1}{2}},
\quad
F_2(\lambda)=-\frac{iy_0^2t\left(\lambda+v_{q,t}-q^2\right)\sinh\left(t\sqrt{\lambda+v_{q,t}}\right)}
{2\sigma^2\Phi_{qt}\left(it\sqrt{\lambda+v_{q,t}}\right)},
\label{E:expo2}
\end{equation}
\begin{equation}
F_3(\lambda)=-\frac{imqy_0t\left(\lambda+v_{q,t}-q^2\right)\left[\cosh\left(t\sqrt{\lambda+v_{q,t}}\right)-1\right]}
{\sigma^2\sqrt{\lambda+v_{q,t}}\Phi_{qt}\left(it\sqrt{\lambda+v_{q,t}}\right)},
\label{E:expo3}
\end{equation}
\begin{equation}
F_4(\lambda)=\frac{im^2q^2t\left(\lambda+v_{q,t}-q^2\right)\left[\sinh\left(t\sqrt{\lambda+v_{q,t}}\right)-t
\sqrt{\lambda+v_{q,t}}\cosh\left(t\sqrt{\lambda+v_{q,t}}\right)\right]}{2\sigma^2\left(\lambda+v_{q,t}\right)
\Phi_{qt}\left(it\sqrt{\lambda+v_{q,t}}\right)},
\label{E:expo4}
\end{equation}
and
\begin{equation}
F_5(\lambda)=\frac{im^2q^3t\left(\lambda+v_{q,t}-q^2\right)\left[4\sinh^2\frac{t\sqrt{\lambda+v_{q,t}}}{2}
-t\sqrt{\lambda+v_{q,t}}\sinh\left(t\sqrt{\lambda+v_{q,t}}\right)\right]}{2\sigma^2\left(\lambda+v_{q,t}\right)
^{\frac{3}{2}}\Phi_{qt}\left(it\sqrt{\lambda+v_{q,t}}\right)}.
\label{E:expo5}
\end{equation}

It is not hard to see that the functions $F_1$, $F_2$, $F_3$, $F_4$, and $F_5$ have removable singularities 
at 
$$
\lambda=-v_{q,t}=\frac{r^2_{qt}}{t^2}.
$$ 
In addition, these functions are analytic in $\mathbb{C}_{+}$. 
Let us denote by $\lambda_1$ the residue of the function $F_1$ at $\lambda=0$. It is not hard to see that
\begin{equation}
\lambda_1=\frac{2r_{qt}^2}{t^2\left|\left(1+qt\right)\cos\left(r_{qt}\right)-r_{qt}\sin\left(r_{qt}\right)\right|}.
\label{E:sim1}
\end{equation}
Our next goal is to apply Theorem \ref{T:lti} to (\ref{E:expo1}). Put 
\begin{equation}
\gamma_1=0,\quad\gamma_2=\frac{1}{2},\quad G_1(\lambda)=\frac{it\sqrt{\lambda+v_{q,t}}}
{\Phi_{qt}\left(it\sqrt{\lambda+v_{q,t}}\right)},
\quad G_2(\lambda)=\frac{\sqrt{2t}}{\sigma}
e^{\frac{qt}{2}}\lambda_1^{\frac{1}{2}},
\label{E:data1}
\end{equation}
and 
\begin{equation}
F(\lambda)=F_2(\lambda)+F_3(\lambda)+F_4(\lambda)+F_5(\lambda).
\label{E:data2}
\end{equation} 
In the sequel, the symbols $\alpha_j$ and $\widetilde{F}_j(0)$
will stand for the numbers in the formulation of Theorem \ref{T:lti} associated with the function $F_j$, $2\le j\le 5$. 
We will next compute these numbers. The following formula will be helpful in the computations:
\begin{equation}
\frac{\left(v_{q,t}-q^2+\lambda\right)\lambda}{\Phi_{qt}\left(it\sqrt{\lambda+v_{q,t}}\right)}
=\frac{\left(v_{q,t}-q^2\right)+\lambda}{\zeta_1+\zeta_2\lambda+\cdots}=\tau_1+\tau_2\lambda+\cdots
\label{E:sim2}
\end{equation}
where 
$$
\zeta_1=\frac{t^2\left(\left(1+qt\right)\cos\left(r_{qt}\right)-r_{qt}\sin\left(r_{qt}\right)\right)}{2r_{qt}},
$$
$$
\zeta_2=\frac{t^2\left(\left(1+qt\right)\cos\left(r_{qt}\right)+r_{qt}\sin\left(r_{qt}\right)\right)}{8r_{qt}^3},
$$
$$
\tau_1=\frac{2\left(v_{q,t}-q^2\right)r_{qt}}{t^2\left(\left(1+qt\right)\cos\left(r_{qt}\right)-r_{qt}\sin\left(r_{qt}\right)\right)}> 0,
$$
and
\begin{align*}
&\tau_2=\frac{\zeta_1-\left(v_{q,t}-q^2\right)\zeta_2}{\zeta_1^2}  \\
&=\frac{4r_{qt}^2
\left(\left(1+qt\right)\cos\left(r_{qt}\right)-r_{qt}\sin\left(r_{qt}\right)\right)-\left(v_{q,t}-q^2\right)
\left(\left(1+qt\right)\cos\left(r_{qt}\right)+r_{qt}\sin\left(r_{qt}\right)\right)}{2r_{qt}t^2
\left(\left(1+qt\right)\cos\left(r_{qt}\right)-r_{qt}\sin\left(r_{qt}\right)\right)^2}.
\end{align*}
Here we use the facts that $v_{q,t}=-\frac{r_{qt}^2}{t^2}< 0$,\,\,\,$\cos\left(r_{qt}\right)< 0$,\,\,\,$\sin\left(r_{qt}\right)> 0$,
\,\,\,$\Phi_{qt}\left(r_{qt}\right)=0$,
$$
\Phi_{qt}^{\prime}\left(r_{qt}\right)=(1+qt)\cos\left(r_{qt}\right)-r_{qt}\sin\left(r_{qt}\right),
$$
and
$$
\Phi_{qt}^{\prime\prime}\left(r_{qt}\right)=-2\sin\left(r_{qt}\right)-\Phi_{qt}\left(r_{qt}\right)=-2\sin\left(r_{qt}\right).
$$

We will next employ (\ref{E:expo2})-(\ref{E:expo5}) to find explicit formulas for the numbers $\alpha_j$ and $\widetilde{F}_j(0)$ 
for \\ $2\le j\le 5$.
It follows from (\ref{E:expo2}), (\ref{E:sim2}), and the formula
$$
\sinh\left(t\sqrt{\lambda+v_{q,t}}\right)=i\sin\left(r_{qt}\right)-i\frac{t^2}{2r_{qt}}\cos\left(r_{qt}\right)\lambda+\cdots
$$
that
\begin{align*}
\lambda F_2(\lambda)&=-\frac{iy_0^2t}{2\sigma^2}\left(\tau_1+\tau_2\lambda+\cdots\right)
\left(i\sin\left(r_{qt}\right)-i\frac{t^2}{2r_{qt}}\cos\left(r_{qt}\right)\lambda+\cdots\right) \\
&=\frac{y_0^2t}{2\sigma^2}\tau_1\sin\left(r_{qt}\right)+\frac{y_0^2t}{2\sigma^2}\left(\tau_2\sin\left(r_{qt}\right)
-\tau_1\frac{t^2}{2r_{qt}}\cos\left(r_{qt}\right)\right)\lambda+\cdots
\end{align*}
Therefore,
\begin{equation}
\alpha_2=\frac{y_0^2t}{2\sigma^2}\tau_1\sin\left(r_{qt}\right)> 0
\label{E:vcr0}
\end{equation}
and
\begin{equation}
\widetilde{F}_2(0)=
\frac{y_0^2t}{2\sigma^2}\tau_1\sin\left(r_{qt}\right)+\frac{y_0^2t}{2\sigma^2}\left(\tau_2\sin\left(r_{qt}\right)
-\tau_1\frac{t^2}{2r_{qt}}\cos\left(r_{qt}\right)\right).
\label{E:vcr1}
\end{equation}

Moreover, (\ref{E:expo3}), (\ref{E:sim2}), and the fact that
$$
\frac{\cosh\left(t\sqrt{\lambda+v_{q,t}}\right)-1}{\sqrt{\lambda+v_{q,t}}}=i\frac{1-\cos\left(r_{qt}\right)}{r_{qt}}
-i\frac{t^3\left(2r_{qt}\sin\left(r_{qt}\right)+\cos\left(r_{qt}\right)-1\right)}{2r_{qt}^3}\lambda+\cdots
$$
give
\begin{align*}
&\lambda F_3(\lambda)=-\frac{im^2q^2t}{2\sigma^2}\left(\tau_1+\tau_2\lambda+\cdots\right)\left(i\frac{1-\cos\left(r_{qt}\right)}{r_{qt}}
-i\frac{t^3\left(2r_{qt}\sin\left(r_{qt}\right)+\cos\left(r_{qt}\right)-1\right)}{2r_{qt}^3}\lambda+\cdots\right) \\
&=\frac{m^2q^2t}{2\sigma^2}\tau_1\frac{1-\cos\left(r_{qt}\right)}{r_{qt}}+\frac{m^2q^2t}{2\sigma^2}
\left[\tau_2\frac{1-\cos\left(r_{qt}\right)}{r_{qt}}-\tau_1
\frac{t^3\left(2r_{qt}\sin\left(r_{qt}\right)+\cos\left(r_{qt}\right)-1\right)}{2r_{qt}^3}\right]\lambda+\cdots
\end{align*}
Hence,
\begin{equation}
\alpha_3=\frac{m^2q^2t}{2\sigma^2}\tau_1\frac{1-\cos\left(r_{qt}\right)}{r_{qt}}> 0
\label{E:vcr2}
\end{equation}
and
\begin{equation}
\widetilde{F}_3(0)=\frac{m^2q^2t}{2\sigma^2}
\left[\tau_2\frac{1-\cos\left(r_{qt}\right)}{r_{qt}}-\tau_1
\frac{t^3\left(2r_{qt}\sin\left(r_{qt}\right)+\cos\left(r_{qt}\right)-1\right)}{2r_{qt}^3}\right].
\label{E:vcr3}
\end{equation}

In addition, (\ref{E:expo4}), (\ref{E:sim2}), and the formulas $r_{qt}\cos\left(r_{qt}\right)+qt\sin\left(r_{qt}\right)=0$ and
\begin{align*}
&\frac{\sinh\left(t\sqrt{\lambda+v_{q,t}}\right)-t\sqrt{\lambda+v_{q,t}}
\cosh\left(t\sqrt{\lambda+v_{q,t}}\right)}{\lambda+v_{q,t}} \\
&=-i\frac{t^2(1+qt)\sin\left(r_{qt}\right)}{r^2_{qt}}-i\frac{t^4\sin\left(r_{qt}\right)(1-qt-r_{qt}^2)}
{r_{qt}^4}\lambda+\cdots
\end{align*}
imply that
\begin{align*}
&\lambda F_4(\lambda)=\frac{im^2q^2t}{2\sigma^2}\left(\tau_1+\tau_2\lambda+\cdots\right)
\left(-i\frac{t^2(1+qt)\sin\left(r_{qt}\right)}{r^2_{qt}}-i\frac{t^4\sin\left(r_{qt}\right)\left(1-qt-r^2_{qt}\right)}
{r_{qt}^4}\lambda+\cdots\right).
\end{align*}
It follows that
\begin{equation}
\alpha_4=\tau_1\frac{m^2q^2t^3}{2\sigma^2}\frac{(1+qt)\sin\left(r_{qt}\right)}{r^2_{qt}}> 0
\label{E:vcr4}
\end{equation}
and
\begin{equation}
\widetilde{F}_4(0)=\frac{m^2q^2t}{2\sigma^2}\left(\tau_1\frac{t^4\sin\left(r_{qt}\right)\left(1-qt-r^2_{qt}\right)}
{r_{qt}^4}+\tau_2\frac{t^2(1+qt)\sin\left(r_{qt}\right)}{r^2_{qt}}\right).
\label{E:vcr5}
\end{equation}

Finally, we have
\begin{align*}
&\frac{4\sinh^2\frac{t\sqrt{\lambda+v_{q,t}}}{2}
-t\sqrt{\lambda+v_{q,t}}\sinh\left(t\sqrt{\lambda+v_{q,t}}\right)}{\left(\lambda+v_{q,t}\right)
^{\frac{3}{2}}}=i\frac{t^3\left(r_{qt}\sin\left(r_{qt}\right)-4\sin^2\frac{r_{qt}}{2}\right)}{r_{qt}^3} \\
&\quad+\left[\frac{i3t^5}{2r_{qt}^5}\left(r_{qt}\sin\left(r_{qt}\right)-4\sin^2\frac{r_{qt}}
{2}\right)+\frac{it^5}{2r_{qt}^3}\left(\frac{3}{r_{qt}}\sin\left(r_{qt}\right)-\cos\left(r_{qt}\right)\right)
\right]\lambda
+\cdots,
\end{align*}
and hence
$$
\lambda F_5(\lambda)=\frac{im^2q^3t}{2\sigma^2}\left(\tau_1+\tau_2\lambda+\cdots\right) 
\left[i\frac{t^3\left(r_{qt}\sin\left(r_{qt}\right)-4\sin^2\frac{r_{qt}}{2}\right)}{r_{qt}^3} 
+D\lambda
+\cdots\right]
$$
where
$$
D=\frac{i3t^5}{2r_{qt}^5}\left(r_{qt}\sin\left(r_{qt}\right)-4\sin^2\frac{r_{qt}}
{2}\right)+\frac{it^5}{2r_{qt}^3}\left(\frac{3}{r_{qt}}\sin\left(r_{qt}\right)-\cos\left(r_{qt}\right)\right).
$$
Using the previous equalities, we obtain
\begin{equation}
\alpha_5=\frac{m^2q^3t^4}{2\sigma^2}\tau_1\frac{4\sin^2\frac{r_{qt}}{2}-r_{qt}\sin\left(r_{qt}\right)}{r_{qt}^3}
\label{E:vcr6}
\end{equation}
and 
\begin{align}
&\widetilde{F}_5(0)=\frac{m^2q^3t^4}{2\sigma^2}\tau_2\frac{4\sin^2\frac{r_{qt}}{2}
-r_{qt}\sin\left(r_{qt}\right)}{r_{qt}^3} \\
&\quad-\frac{m^2q^3t^4}{2\sigma^2}\tau_1\left[\frac{3t^5}{2r_{qt}^5}\left(r_{qt}
\sin\left(r_{qt}\right)-4\sin^2\frac{r_{qt}}
{2}\right)+\frac{t^5}{2r_{qt}^3}\left(\frac{3}{r_{qt}}\sin\left(r_{qt}\right)-\cos\left(r_{qt}\right)\right)\right].
\label{E:vcr7}
\end{align}
Our next goal is to prove that $\alpha_5> 0$. Indeed, we have
\begin{equation}
4\sin^2\frac{r_{qt}}{2}-r_{qt}\sin\left(r_{qt}\right)=2\sin\frac{r_{qt}}{2}\left(2\sin\frac{r_{qt}}{2}
-r_{qt}\cos\frac{r_{qt}}{2}
\right).
\label{E:chto}
\end{equation}
Since $\frac{\pi}{4}<\frac{r_{qt}}{2}<\frac{\pi}{2}$ 
and $x<\tan x$ for $\frac{\pi}{4}< x<\frac{\pi}{2}$,
inequality $\alpha_5> 0$ follows from (\ref{E:vcr6}) and (\ref{E:chto}).

We can now apply Theorem \ref{T:lti} with the data given by (\ref{E:data1}) and (\ref{E:data2}). Recall that 
we denoted by $\alpha$ the positive number given by
\begin{equation}
\alpha=\alpha_2+\alpha_3+\alpha_4+\alpha_5
\label{E:chi1}
\end{equation}
where $\alpha_2$, $\alpha_3$, $\alpha_4$, and $\alpha_5$ are defined in (\ref{E:vcr0}), (\ref{E:vcr2}), (\ref{E:vcr4}), 
and (\ref{E:vcr6}), respectively. We also denoted by $F$ the function $F_2+F_3+F_4+F_5$ (see (\ref{E:data2})), and put
\begin{equation}
\widetilde{F}(0)=\widetilde{F}_2(0)+\widetilde{F}_3(0)+\widetilde{F}_4(0)+\widetilde{F}_5(0)
\label{E:chi2}
\end{equation}
where $\widetilde{F}_2(0)$, $\widetilde{F}_3(0)$, $\widetilde{F}_4(0)$, and $\widetilde{F}_5(0)$ are given by 
(\ref{E:vcr1}), (\ref{E:vcr3}), (\ref{E:vcr5}), and (\ref{E:vcr7}), respectively. 
By applying Theorem \ref{T:lti} to (\ref{E:expo1}), we see that
\begin{equation}
y^{-\frac{1}{2}}\exp\left\{\left(q^2-v_{q,t}\right)y\right\}
m_t\left(\frac{\sqrt{2}\sigma}{\sqrt{t}}\sqrt{y}\right)
=\frac{\sqrt{t}}{\sqrt{2\pi}\sigma}
e^{\frac{qt}{2}}\lambda_1^{\frac{1}{2}}e^{\widetilde{F}(0)}y^{-\frac{1}{2}}
e^{2\sqrt{\alpha}\sqrt{y}}\left(1+O\left(y^{-\frac{1}{4}}\right)\right)
\label{E:ou2}
\end{equation}
as $y\rightarrow\infty$. Next, replacing $y$ by $y^2\frac{t}{2\sigma^2}$ in formula (\ref{E:ou2}), we obtain
\begin{align}
m_t\left(y\right)&=\frac{\sqrt{t}}{\sqrt{2\pi}\sigma}
e^{\frac{qt}{2}}\lambda_1^{\frac{1}{2}}e^{\widetilde{F}(0)}
\exp\left\{\frac{\sqrt{2\alpha t}}{\sigma}y\right\}\exp\left\{-\frac{t\left(q^2-v_{q,t}\right)}{2\sigma^2}y^2\right\}
\left(1+O\left(y^{-\frac{1}{2}}\right)\right)
\label{E:ou3}
\end{align}
as $y\rightarrow\infty$. Now, it is clear that formula (\ref{E:ou3}) implies Theorem \ref{T:ohh}. Moreover, the following lemma holds:
\begin{lemma}\label{L:con3}
The constants $E$, $F$, and $G$ in Theorem \ref{T:ohh}
are given by 
$$
E=\frac{\sqrt{t}}{\sqrt{2\pi}\sigma}
e^{\frac{qt}{2}}\lambda_1^{\frac{1}{2}}e^{\widetilde{F}(0)},\quad
F=\sigma^{-1}\sqrt{2\alpha t},\quad\mbox{and}\quad
G=\left(2\sigma^2\right)^{-1}t\left(q^2+t^{-2}r_{qt}^2\right),
$$
where the numbers $\lambda_1> 0$, $\alpha> 0$, and $\widetilde{F}(0)$ are defined 
in (\ref{E:sim1}), (\ref{E:chi1}), and (\ref{E:chi2}), respectively.
\end{lemma}

This completes the proof of Theorem \ref{T:ohh}.
\section{Proof of Theorem \ref{T:maino}}\label{S:qv}
We will first show that it is possible to apply Theorem \ref{T:into} with
$A(y)=y^{-1}m_t(y)e^{Gy^2}$, $k=\sqrt{G+\frac{t}{8}}$, $l=F$, $\zeta(y)=Ey^{-1}$, and $b(y)=y^{-\frac{1}{2}}$,
where the constants $E$, $F$, and $G$ are such as in Lemma \ref{L:con3}.
It is not hard to see that condition 2 in Theorem \ref{T:into} follows from formula (\ref{E:ohh1}). 
In addition, it is clear that condition 3 with $\gamma=1$ holds. The validity of condition 1 in Theorem \ref{T:into} can be shown by reasoning as in
the proof of Lemma \ref{L:dodo}. Here we use formula (\ref{E:expo1}) instead of
formula (\ref{E:laai}).

It follows from Theorem \ref{T:into} that
\begin{align}
&\int_0^{\infty}y^{-1}m_t(y)\exp\left\{-\left(\frac{z^2}{y^2}+\frac{ty^2}{8}\right)\right\}dy \nonumber \\
&=E\frac{\sqrt{\pi}}{2k}\exp\left\{\frac{F^2}{16k^2}\right\}k^{\frac{1}{2}}z^{-\frac{1}{2}}
\exp\left\{Fk^{-\frac{1}{2}}\sqrt{z}\right\}e^{-2kz}\left(1+O\left(z^{-\frac{1}{4}}\right)\right)
\label{E:do2}
\end{align}
as $z\rightarrow\infty$. Replacing $z$ by 
$\displaystyle{\frac{\log x}{\sqrt{2t}}}$ in 
formula (\ref{E:do2}) and taking into account formula (\ref{E:stvoll1}) and the equality 
$\displaystyle{k=\frac{\sqrt{8G+t}}{2\sqrt{2}}}$, we obtain
\begin{align}
&D_t\left(x_0e^{\mu t}x\right)=\frac{E}{x_0e^{\mu t}}2^{-\frac{1}{2}}t^{-\frac{1}{4}}
(8G+t)^{-\frac{1}{4}}\exp\left\{\frac{F^2}{2(8G+t)}\right\}
 \nonumber \\
&\quad(\log x)^{-\frac{1}{2}}
\exp\left\{\frac{F\sqrt{2}}{t^{\frac{1}{4}}(8G+t)^{\frac{1}{4}}}\sqrt{\log x}\right\}x^{-\left(\frac{3}{2}+
\frac{\sqrt{8G+t}}{2\sqrt{t}}\right)}
\left(1+O\left((\log x)^{-\frac{1}{4}}\right)\right)
\label{E:do3}
\end{align}
as $x\rightarrow\infty$.

Now, it is clear that formula (\ref{E:do3}) implies Theorem \ref{T:maino}. In addition, the following lemma holds:
\begin{lemma}\label{L:con4}
The constants $B_1$, $B_2$, and $B_3$ 
are given by 
$B_1=\frac{E}{x_0e^{\mu t}}2^{-\frac{1}{2}}t^{-\frac{1}{4}}
(8G+t)^{-\frac{1}{4}}\exp\left\{\frac{F^2}{2(8G+t)}\right\}$,
$B_2=\frac{F\sqrt{2}}{t^{\frac{1}{4}}(8G+t)^{\frac{1}{4}}}$, and
$B_3=\frac{3}{2}+
\frac{\sqrt{8G+t}}{2\sqrt{t}}$,
where the numbers $E$, $F$, and $G$ are defined 
in Lemma \ref{L:con3}.
\end{lemma} 

It is an interesting fact that Theorem \ref{T:maino} with $m=0$ is a special case of Theorem \ref{T:main}. 
This will be explained below. Recall that in model (\ref{E:SS}),
the volatility process is the absolute value of an Ornstein-Uhlenbeck process.
The following explicit representation is valid for the Ornstein-Uhlenbeck process $\widetilde{Y}_t$ satisfying the 
stochastic differential equation 
$d\widetilde{Y}_t=q\left(m-\widetilde{Y}_t\right)dt+\sigma dZ_t$:
$$
\widetilde{Y}_t\left(q,m,\sigma,y_0\right)=e^{-qt}y_0+\left(1-e^{-qt}\right)m+\sigma e^{-qt}\int_0^te^{qu}dZ_u
$$
(see, e.g., \cite{keyN}, Proposition 3.8). Therefore,
\begin{equation}
\widetilde{Y}_t\left(q,m,\sigma,y_0\right)=\widetilde{Y}_t\left(q,0,\sigma,y_0+\left(e^{qt}-1\right)m\right).
\label{E:ss2}
\end{equation}
It is known that squared Ornstein-Uhlenbeck processes are related to CIR-processes. Indeed, 
it is not hard to see, using the It\^{o} formula, that the squared Ornstein-Uhlenbeck process 
$T_t=\widetilde{Y}_t\left(q,0,\sigma,y_0\right)^2$ satisfies the following 
stochastic differential equation:
$dT_t=\left(\sigma^2-2qT_tdt\right)+2\sigma\sqrt{T_t}dZ_t$.
Therefore, the uniqueness implies that the process $\widetilde{Y}_t\left(q,0,\sigma,z_0\right)^2$ is indistinguishable from the CIR-process
$Y_t\left(\sigma^2,-2q,2\sigma,z_0^2\right)$. It follows from (\ref{E:ss2}) that
\begin{equation}
\widetilde{Y}_t\left(q,m,\sigma,y_0\right)^2=Y_t\left(\sigma^2,-2q,2\sigma,\left(y_0+\left(e^{qt}-1\right)m\right)^2\right),
\label{E:beou1}
\end{equation}
and hence in the case where $m=0$, the mixing distribution densities corresponding to the processes on the both 
sides of (\ref{E:beou1}) coincide. Therefore, all the results concerning the distribution density of the stock price 
process in model (\ref{E:H}) can be reformulated for model (\ref{E:SS}) with $m=0$. For instance, formula 
(\ref{E:lai2}) becomes
\begin{align*}
&\int_0^{\infty}\exp\left\{-\lambda y\right\}y^{-\frac{1}{2}}
m_t\left(t^{-\frac{1}{2}}y^{\frac{1}{2}};q,0,\sigma,y_0\right)dy \\
&=2\sqrt{t}\exp\left\{\frac{qt}{2}\right\}\left(\frac{\sqrt{q^2+2\sigma^2\lambda}}
{\sqrt{q^2+2\sigma^2\lambda}\cosh(t\sqrt{q^2+2\sigma^2\lambda})
+q\sinh(t\sqrt{q^2+2\sigma^2\lambda})}\right)^{\frac{1}{2}} \\
&\quad\exp\left\{-\frac{y_0^2\lambda\sinh(t\sqrt{q^2+2\sigma^2\lambda})}
{\sqrt{q^2+2\sigma^2\lambda}\cosh(t\sqrt{q^2+2\sigma^2\lambda})
+q\sinh(t\sqrt{q^2+2\sigma^2\lambda})}\right\}.
\end{align*}
Summarizing what was said above, we see that Theorem \ref{T:maino} with $m=0$ follows from Theorem \ref{T:main} and formula (\ref{E:beou1}).

The next statement concerns the moment explosion problem for the Stein-Stein model.
\begin{lemma}\label{L:blupo}
Let $q\ge 0$, $m\ge 0$, $\sigma> 0$, and $p\in\mathbb{R}$. Then, the following statement is true for the moment $l_p$ 
of the stock price distribution density $D_t$ in model (\ref{E:SS}): 
$$
l_p<\infty\Longleftrightarrow\frac{1}{2}-\frac{\sqrt{8C+t}}{2\sqrt{t}}< p< \frac{1}{2}+\frac{\sqrt{8C+t}}{2\sqrt{t}},
$$ 
where $\displaystyle{C=\frac{1}{2\sigma^2}\left(tq^2+t^{-1}r_{tq}^2\right)}$.
\end{lemma}

It is not hard to see that Lemma \ref{L:blupo} follows from Lemma \ref{L:blup} and formula (\ref{E:beou1}).
\section{Proof of Theorem \ref{T:impik}}
Let $\bar{K}> 0$, and let $f$ and $g$ be positive functions on the interval 
$[\bar{K},\infty)$. We will write 
$f(x)\approx g(x)$, $x\rightarrow\infty$, if there exist constants $c_1> 0$, $c_2> 0$, 
and $K_0>\bar{K}$ such that the inequalities
$c_1g(x)\le f(x)\le c_2g(x)$ hold for all $x> K_0$. The following notation will be used below: 
$\hat{V}_0(k)=V_0(K)$ (see (\ref{E:opt}) for the definition of $V_0$). 

The next lemma was obtained in \cite{keyGS3}.
\begin{lemma}\label{L:genet}
Suppose that there exist positive increasing continuous functions $\psi$ and $\phi$ 
such that \\
$\lim_{k\rightarrow\infty}\psi(k)=\lim_{k\rightarrow\infty}\phi(k)=\infty$
and
\begin{equation}
\hat{V}_0(k)\approx\frac{\psi(k)}{\phi(k)}\exp\left\{-\frac{\phi(k)^2}{2}\right\}
\label{E:gene30}
\end{equation}
as $K\rightarrow\infty$. Then the following asymptotic formula holds:
$$
\hat{I}(k)=\frac{1}{\sqrt{T}}\left(\sqrt{2k+\phi(k)^2}-\phi(k)\right)
+O\left(\frac{\psi(k)}{\phi(k)}\right)
$$ 
as $K\rightarrow\infty$.
\end{lemma}

For the model in (\ref{E:H}), let $\psi$ be a function such as in the formulation of Lemma \ref{L:genet}. 
Our next goal is to find a function $\phi$ for which formula (\ref{E:gene30}) holds.
It follows from (\ref{E:option}) that
\begin{equation}
\hat{V}_0(k)=e^{-rT}\left(x_0e^{rT}\right)^2\left[\int_{e^k}yD_T\left(x_0e^{rT}y\right)dy-e^k\int_{e^k}^{\infty}
D_T\left(x_0e^{rT}y\right)dy\right].
\label{E:zza1}
\end{equation}
Now it is not hard to see that (\ref{E:dopo}) and (\ref{E:zza1}) imply
\begin{equation}
\hat{V}_0(k)\approx k^{-\frac{3}{4}+\frac{a}{c^2}}e^{A_2\sqrt{k}}e^{-\left(A_3-2\right)k},\quad k\rightarrow\infty.
\label{E:zza2}
\end{equation}
Put 
$$
\phi(k)=\sqrt{\left(2A_3-4\right)k-2A_2\sqrt{k}
+\left(\frac{1}{2}-\frac{2a}{c^2}\right)\log k-2\log\psi(k)}.
$$
Then we have $\phi(k)\approx\sqrt{k}$, and (\ref{E:zza2}) shows that condition (\ref{E:gene30}) in Lemma \ref{L:genet}
holds. Applying this lemma and the mean value theorem, we see that 
\begin{align*}
\hat{I}(k)&=\frac{\sqrt{2}}{\sqrt{T}}\left[\sqrt{\left(A_3-1\right)k
-A_2\sqrt{k}+\left(\frac{1}{4}-\frac{a}{c^2}\right)\log k}-\sqrt{\left(A_3-2\right)k-A_2\sqrt{k}+\left(\frac{1}{4}
-\frac{a}{c^2}\right)\log k}\right] \\
&+O\left(\frac{\psi(k)}{\sqrt{k}}\right),\quad k\rightarrow\infty.
\end{align*}

Next using the fact that $\sqrt{1-h}=1-\frac{1}{2}h+O\left(h^2\right)$ as $h\downarrow 0$, we obtain (\ref{E:impik1}). The proof of 
(\ref{E:impik2}) is similar. Here we use (\ref{E:dop1}) instead of (\ref{E:dopo}).


\begin{thebibliography}{11} 
\bibitem[1]{keyAP} Andersen, L. B. G., Piterbarg, V. V.: {Moment explosions in stochastic volatility models.} Finance Stoch. \bf 11,
\rm 29-50 (2007)
\bibitem[2]{keyBF} Benaim, S., Friz, P.: {Regular variation and smile asymptotics.} To appear in 
Math. Finance. 
\bibitem[3]{keyBFL} Benaim, S., Friz, P., Lee, R.: {The Black-Scholes implied volatility at extreme strikes.} 
Preprint (2007) 
\bibitem[4]{keyBS} Borodin, A. N., Salminen, P.: {Handbook of Brownian Motion - Facts and Formulae.}
Birkh\"{a}user Verlag, Basel (1996)
\bibitem[5]{keyCS} Carr, P., Schr\"{o}der, M.: {Bessel processes, the integral of Geometric Brownian motion, and 
Asian options.} Theory Probab. Appl. \bf 48, \rm 400-425 (2004)
\bibitem[6]{keyCIR} Cox, J. C., Ingersoll, J. E., Ross, S. A.: {A theory of the term structure of interest rates.} 
Econometrica \bf 53, \rm 385-407 (1985)
\bibitem[7]{keyDY} Dr$\rm\breve{a}$gulescu, A. A., Yakovenko, V. M.: {Probability distribution of returns in the Heston model with stochastic volatility.} Quantitative Finance \bf 2, \rm 443-453 (2002)
\bibitem[8]{keyD} Dufresne, D.: {Bessel processes and a functional of Brownian motion.}
University of Melbourne ePrints Repository (2004)
\bibitem[9]{keyFPS} Fouque, J.-P., Papanicolaou, G., Sircar, K. R.: {Derivatives in Financial Markets with Stochastic Volatility.} Cambridge University Press, Cambridge (2000)
\bibitem[10]{keyGY} G\"{o}ing-Yaeschke, A., Yor, M.: {A survey and some generalizations of Bessel processes.} Bernoulli 
\bf 9, \rm 313-349 (2003)
\bibitem[11]{keyGS1} Gulisashvili, A., Stein, E. M.:
{Asymptotic behavior of the distribution of the stock price in models with stochastic volatility: 
the Hull-White model.} C. R. Acad Sci. Paris, Ser. I \bf 343, \rm 519-523 (2006)
\bibitem[12]{keyGS2} Gulisashvili, A., Stein, E. M.: {Asymptotic behavior of distribution densities in models with stochastic volatility, I.} Submitted for publication.
\bibitem[13]{keyGS3} Gulisashvili, A., Stein, E. M.: {Implied volatility in the Hull-White model.} To be published in Mathematical Finance.
\bibitem[14]{keyH} Heston, S. L.: {A closed-form solution for options with stochastic volatility, 
with applications to bond and currency options.} Review of Financial Studies \bf 6, \rm 327-343 (1993) 
\bibitem[15]{keyL} Lee, R.: {The moment formula for implied volatility at extreme strikes.} 
Math. Finance \bf 14, \rm 469-480 (2004)
\bibitem[16]{keyN} Nielsen, L. T.: {Pricing and Hedging of Derivative Securities.} 
Oxford University Press, Oxford (1999)
\bibitem[17]{keyPY} Pitman, J., Yor, M.: {A decomposition of Bessel bridges.} Z. 
Wahrscheinlichkeitstheorie verw. Gebiete \bf 59, \rm 425-457 (1982)
\bibitem[18]{keyRY} Revuz, D., Yor, M.: {Continuous Martingales and Brownian motion.}
Springer-Verlag, Berlin (1991)
\bibitem[19]{keySS1} Stein, E. M., Stein, J.: {Stock price distributions with stochastic volatility:
An analytic approach.} Review of Financial Studies \bf 4, \rm 727-752 (1991)
\bibitem[20]{keySS2} Stein, E. M., Shakarchi, R.: {Complex Analysis.}
Princeton University Press, Princeton and Oxford (2003)
\bibitem[21]{keyW} Wenocur, M. L.: {Ornstein-Uhlenbeck process with quadratic killing,} Journal of Applied Probability 
\bf 27, \rm 707-712 (1990)
\end{thebibliography}
\end{document}